\documentclass[sigconf,nonacm]{acmart}
\AtBeginDocument{%
  \providecommand\BibTeX{{%
    \normalfont B\kern-0.5em{\scshape i\kern-0.25em b}\kern-0.8em\TeX}}}

\usepackage{subcaption}
\graphicspath{{Figures/}}
\usepackage{multirow}
\usepackage{multicol}
\usepackage[noend]{algpseudocode}
\usepackage{algorithm}

\DeclareMathOperator*{\argmin}{argmin}

\begin{document}

\title{Spatial Parquet: A Column File Format for Geospatial Data Lakes [Extended Version]}\titlenote{a short version of this paper is published in SIGSPATIAL'22 at \url{https://doi.org/10.1145/3557915}}

\author{Majid Saeedan}
\email{msaee007@ucr.edu}
\affiliation{%
  \institution{University of California, Riverside}
  \streetaddress{900 University Ave}
  \city{Riverside}
  \state{California}
  \country{USA}
  \postcode{92521}
}
\author{Ahmed Eldawy}
\email{eldawy@ucr.edu}
\affiliation{%
  \institution{University of California, Riverside}
  \streetaddress{900 University Ave}
  \city{Riverside}
  \state{California}
  \country{USA}
  \postcode{92521}
}


\begin{abstract}
Modern data analytics applications prefer to use column-storage formats due to their improved storage efficiency through encoding and compression. Parquet is the most popular file format for column data storage that provides several of these benefits out of the box. However, geospatial data is not readily supported by Parquet.
This paper introduces Spatial Parquet, a Parquet extension that efficiently supports geospatial data. Spatial Parquet inherits all the advantages of Parquet for non-spatial data, such as rich data types, compression, and column/row filtering. Additionally, it adds three new features to accommodate geospatial data. First, it introduces a geospatial data type that can encode all standard spatial data types in a column format compatible with Parquet. Second, it adds a new lossless and efficient encoding method, termed FP-delta, that is customized to efficiently store geospatial coordinates stored in floating-point format. Third, it adds a light-weight spatial index that allows the reader to skip non-relevant parts of the file for increased read efficiency. Experiments on large-scale real data showed that SpatialParquet can reduce the data size by a factor of three even without compression. Compression can further reduce the storage size. Additionally, Spatial Parquet can reduce the reading time by two orders of magnitude when the light-weight index is applied. This initial prototype can open new research directions to further improve geospatial data storage in column format.
\end{abstract}





\maketitle

\section{Introduction}

Recently, there has been a tremendous increase in the amount of publicly available data that are used for data science and data analysis projects. For example, Data.gov~\cite{datagov} contains more than 350,000 dataset that are provided by the US federal government alone. Other governmental and non-governmental open data repositories provide non-precedented volumes of data that grow faster than Moore's Law. These dataset open the door for many interesting data science projects but maintaining all this data is challenging. These dataset are often kept in data warehouses or {\em data lakes} where users can browse, download, and analyze all this data.

To store any dataset on disk, the two major formats are row-oriented and column-oriented formats. Traditional row-oriented formats, such as CSV and JSON, store the entire record in consecutive disk locations. These formats are usually easier to process and are suitable when the entire record is needed. However, for analytical jobs that need to access a few fields, i.e., columns, it adds unnecessary overhead. Thus, column-oriented formats have been proposed to overcome these limitations. In column formats, the entire column is stored in consecutive bytes on disk which provides two unique advantages over row formats. First, if only a few columns are needed for an analytical job, e.g., calculate average income, we can scan this entire column while not reading the rest of the file from disk. Second, it enables more efficient encoding techniques, such as delta encoding, to store each column in a more efficient way. In summary, column formats are preferred for large scale analytical queries.

One of the most popular column formats is Parquet~\cite{V16} which is an open-source file format inspired by Google's Dremel~\cite{MGL+10} system. Parquet is more geared towards big variety data by allowing nested and repeated attributes such as in JSON files. Similar to other column formats, it supports a library of encoding and compression techniques for numeric values to increase its efficiency.

With the increasing amount of geospatial data, Parquet is a very attractive solution that has the potential of saving a significant amount of disk space while increasing the performance of data analysis jobs. However, Parquet is not readily suitable for geospatial data that is more complicated than simple numeric values. In particular, Parquet has three main limitations that limit its use with geospatial data. First, geospatial data is stored as points, lines, and polygons, which have some internal structure that Parquet does not understand. Second, geospatial data consists mainly of $(x,y)$ coordinates that are stored in floating-point format but Parquet does not provide an efficient encoder for floating-point values. Third, Parquet provides the feature of column statistics that can be used as an index but it does not work for geospatial data. The only solution that is currently available is GeoParquet~\cite{geoparquet} which partially addresses the first challenge and adds a significant overhead which defies the purpose of using a column store in the first place.

To resolve these issues, this paper presents SpatialParquet, an extension to the Parquet file format that overcomes the limitations of Parquet. First, it proposes a new data type that is compatible with the Parquet file format and can store all common geometry types, i.e., Point, LineString, Polygon, MultiPoint, MultiLineString, and MultiPolygon. Second, it adds a novel FP-delta encoder that can significantly reduce the storage requirements for floating-point values that represent geospatial coordinates. Third, it combines the statistics feature of Parquet with the proposed structure to provide a light-weight spatial index that can skip disk pages that do not match a given spatial query range. We run extensive experimental evaluation and found that SpatialParquet outperforms all existing file formats in terms of storage size. 

The rest of this paper is organized as follows. Section~\ref{sec:structure} explains how we structure all standard geospatial data in Spatial Parquet. Section~\ref{sec:encoding} describes the FP-delta encoding method for floating-point geospatial coordinates. Section~\ref{sec:indexing} introduces the light-weight spatial index. Experimental evaluation results are detailed in Section~\ref{sec:experiments}. The related work is explained in Section~\ref{sec:related-work}. Finally, Section~\ref{sec:conclusion} concludes the paper.

\section{The Structure}
\label{sec:structure}

This section describes how SpatialParquet stores all the geometry attributes into a unified structure when writing to disk and how it reconstructs them when reading back from disk. There are two main challenges that SpatialParquet has to overcome. First, since Parquet requires all records to have the same schema, we need to create one common schema that can support all the geometry types, i.e., Point, LineString, Polygon, MultiPoint, MultiLineString, and MultiPolygon. GeometryCollection requires special handling that we mention at the end. The second challenge is to ensure that this structure keeps the semantic meaning of all the individual parts of the geometries to facilitate efficient storage and retrieval, e.g., the coordinates and sub-parts of some geometries.

To overcome the two challenges above, we propose the following schema to store geometries. We use Google Protocol Buffers Format (PBF) which is the one used by Parquet.

\begin{verbatim}
message Geometry {
  required int type;
  repeated group part {
    repeated group coordinate {
      required double x;
      required double y;
    }
  }
}
\end{verbatim}

Now, let us explain the structure above. The {\tt type} attribute stores a numerical value that represents the geometry type, i.e., 1=Point, 2=LineString, ... etc. We reserve type 0 to represent empty geometries. The outer group, {\tt part}, represents a connected component in the geometry. For example, in a Polygon, the outer shell and each inner hold is a part. Finally, the {\tt coordinate} group represents a sequence of coordinates that comprise one part. For brevity, this paper assumes two-dimensional coordinates but the structure above can be directly extended to support three dimensions or more by adding their values in the inner-most group. Notice that PBF allows any level of nesting so if the geometry is a part of a feature along with other attributes, the entire definition above will be a single attribute in the feature as shown below where the `...' will be replaced by the definition above.

\begin{verbatim}
message Feature {
  required int id;
  optional string name;
  // Other non-spatial attributes
  optional group geometry { ... }
}
\end{verbatim}

Now, looking at the structure above, we can see that it overcomes the two challenges described earlier. First, this unified structure can support all geometry types as detailed later in this section. Second, this structure contains three columns, {\tt type}, {\tt x}, and {\tt y}, where each one holds a semantic meaning to the geometry and all of them are visible to Parquet to store them efficiently. Additionally, the overhead of maintaining the double nested group, i.e., {\tt part} and {\tt coordinate}, is minimal thanks to the Parquet structure. In this specific case, only four extra bits are needed for the {\tt x} and {\tt y} attributes, two bits for the definition level and tow bits for the repetition level. Interested readers can refer to the Dremel paper~\cite{MGL+10} for more details about the definition and repetition levels.

\subsection{Point (type=1)}

\begin{figure}
    \centering
    \begin{subfigure}{0.35\linewidth}
    \includegraphics[width=\linewidth]{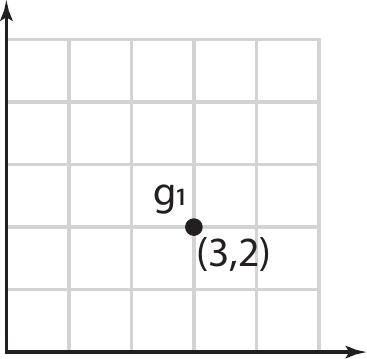}
    \caption{Sample point}
    \end{subfigure}
    \begin{subfigure}{0.49\linewidth}
    \small
        \begin{tabular}{|c|r|r|}
        \hline
         \multirow{3}{*}{Type} & \multicolumn{2}{l|}{Part} \\ \cline{2-3}
         & \multicolumn{2}{l|}{Coordinate} \\ \cline{2-3}
         & $x$ & $y$ \\ \hline \hline
         1 & 3 & 2 \\ \hline
        \end{tabular}
    \caption{Column Representation}
    \end{subfigure}
    \caption{Column representation for a Point}
    \label{fig:geometry-point}
\end{figure}

A Point contains a single coordinate $(x,y)$ which can be represented as shown in Figure~\ref{fig:geometry-point}.

\subsection{LineString (type=2)}

\begin{figure}
    \centering
    \begin{subfigure}{0.35\linewidth}
        \includegraphics[width=\linewidth]{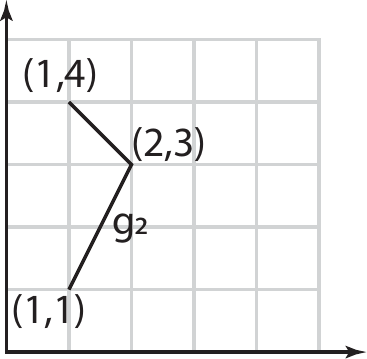}
        \caption{Sample LineString}
    \end{subfigure}
        \begin{subfigure}{0.49\linewidth}
        \small
        \begin{tabular}{|c|r|r|}
        \hline
         \multirow{3}{*}{Type} & \multicolumn{2}{l|}{Part} \\ \cline{2-3}
         & \multicolumn{2}{l|}{Coordinate} \\ \cline{2-3}
         & $x$ & $y$ \\ \hline \hline
         2 & 1 & 1 \\
         & 2 & 3 \\
         & 1 & 4 \\ \hline
        \end{tabular}
        \caption{Column Representation}
    \end{subfigure}
    \caption{Column representation for a LineString}
    \label{fig:geometry-linestring}
\end{figure}

A LineString is a sequence of coordinates $\langle (x_1,y_1), (x_2, y_2)$, $\ldots$, $(x_n,y_n)\rangle$. Figure~\ref{fig:geometry-linestring} gives an example of a LineString, $g_2$, with three points. The points are all represented within one {\tt part} in their respective order.

\subsection{Polygon (type=3)}

\begin{figure}
    \centering
    \begin{subfigure}{0.35\linewidth}
        \includegraphics[width=\linewidth]{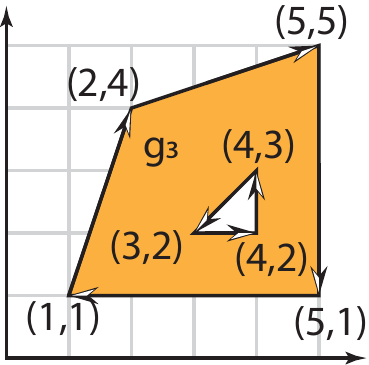}
        \caption{Sample Polygon}
    \end{subfigure}
        \begin{subfigure}{0.49\linewidth}
        \footnotesize
        \begin{tabular}{|c|r|r|}
        \hline
         \multirow{3}{*}{Type} & \multicolumn{2}{l|}{Part} \\ \cline{2-3}
         & \multicolumn{2}{l|}{Coordinate} \\ \cline{2-3}
         & $x$ & $y$ \\ \hline \hline
         3 & 1 & 1 \\
           & 2 & 4 \\
           & 5 & 5 \\
           & 5 & 1 \\
           & 1 & 1 \\ \cline{2-3}
           & 3 & 2 \\
           & 4 & 2 \\
           & 4 & 3 \\ \hline
        \end{tabular}
        \caption{Column Representation}
    \end{subfigure}
    \caption{Column representation for a Polygon}
    \label{fig:geometry-polygon}
\end{figure}

A Polygon contains a list of {\em rings}. Each ring is a LineString that has the same starting and ending points, i.e., $(x_1,y_1)=(x_n,y_n)$. The first ring represents the outer shell while the subsequent rings represent the inner holes. In SpatialParquet, each ring is represented as a {\tt part} similar to how we store a LineString. For consistency, we follow a common convention for storing polygons where the outer shell is stored in clock-wise (CW) order while inner holes are stored in counter clock-wise (CCW) order. Notice that we do not need this information when parsing a polygon since there is only one outer shell, i.e., the first ring, and all subsequent rings are holes. However, this information will become useful for MultiPolygons.

Figure~\ref{fig:geometry-polygon} illustrates an example of a polygon, $g_3$, with one hole. The outer shell contains four segments that are stored in CW order. Notice that we repeat the last point which is similar to the first point as a common convention in most spatial file formats even though it could be redundant.  The inner hole is stored as a second part and the points are ordered in CCW order. In the table representation, we use a horizontal line to represent the end of each part. In Parquet, this is represented by using a repetition level = 1 for the first value in the second ring.

\subsection{MultiPoint (type=4)}

\begin{figure}
    \centering
    \begin{subfigure}{0.35\linewidth}
        \includegraphics[width=\linewidth]{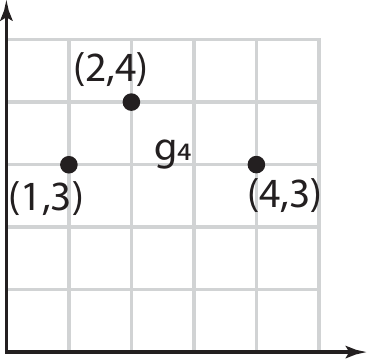}
        \caption{Sample MultiPoint}
    \end{subfigure}
        \begin{subfigure}{0.49\linewidth}
        \small
        \begin{tabular}{|c|r|r|}
        \hline
         \multirow{3}{*}{Type} & \multicolumn{2}{l|}{Part} \\ \cline{2-3}
         & \multicolumn{2}{l|}{Coordinate} \\ \cline{2-3}
         & $x$ & $y$ \\ \hline \hline
         4 & 1 & 3 \\ \cline{2-3}
           & 2 & 4 \\ \cline{2-3}
           & 4 & 3 \\ \hline
        \end{tabular}
        \caption{Column Representation}
    \end{subfigure}
    \caption{Column representation for a MultiPoint}
    \label{fig:geometry-multipoint}
\end{figure}

A MultiPoint consists of a sequence of independent point locations. Each point is represented as a single coordinate $(x,y)$. A MultiPoint is represented in SpatialParquet by creating a separate part for each point with a single coordinate inside it. Figure~\ref{fig:geometry-multipoint} shows an example of a single MultiPoint, $g_4$, with three points inside it. Notice how each point is stored as a separate part. We could also store all the points in one part and it will use exactly the same storage size. However, we chose to use a single part for each point as it is semantically more accurate.

\subsection{MultiLineString (type=5)}

\begin{figure}
    \centering
    \begin{subfigure}{0.49\linewidth}
        \includegraphics[width=0.71\linewidth]{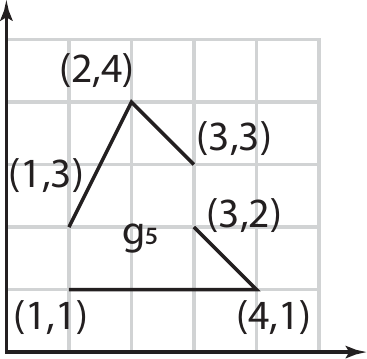}
        \caption{Sample MultiLineString}
    \end{subfigure}
        \begin{subfigure}{0.49\linewidth}
        \small
        \begin{tabular}{|c|r|r|}
        \hline
         \multirow{3}{*}{Type} & \multicolumn{2}{l|}{Part} \\ \cline{2-3}
         & \multicolumn{2}{l|}{Coordinate} \\ \cline{2-3}
         & $x$ & $y$ \\ \hline \hline
         5 & 1 & 1 \\
           & 4 & 1 \\
           & 3 & 2 \\ \cline{2-3}
           & 3 & 3 \\
           & 2 & 4 \\
           & 1 & 3 \\ \hline
        \end{tabular}
        \caption{Column Representation}
    \end{subfigure}
    \caption{Column representation for a MultiLineString}
    \label{fig:geometry-multilinestring}
\end{figure}

A MultiLineString consists of multiple line strings. Each line string is a sequence of coordinates. SpatialParquet stores MultiLineStrings by creating a separate part for each LineString. Each part contains the sequence of coordinates as done with LineString. Figure~\ref{fig:geometry-multilinestring} illustrates an example with a MultiLineString, $g_5$, that contains two LineStrings. Each LineString is represented as a separate part in the column representation.

\subsection{MultiPolygon (type=6)}

\begin{figure}
    \centering
    \begin{subfigure}{0.49\linewidth}
        \includegraphics[width=0.71\linewidth]{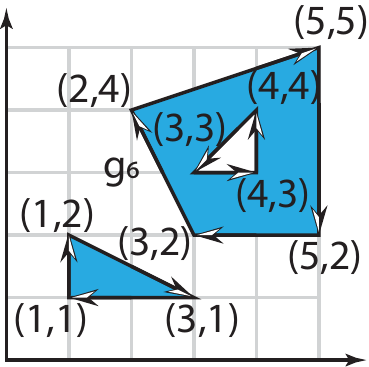}
        \caption{Sample MultiPolygon}
    \end{subfigure}
        \begin{subfigure}{0.49\linewidth}
        \footnotesize
        \begin{tabular}{|c|r|r|}
        \hline
         \multirow{3}{*}{Type} & \multicolumn{2}{l|}{Part} \\ \cline{2-3}
         & \multicolumn{2}{l|}{Coordinate} \\ \cline{2-3}
         & $x$ & $y$ \\ \hline \hline
         6 & 2 & 4 \\
           & 5 & 5 \\
           & 5 & 2 \\
           & 3 & 2 \\
           & 2 & 4 \\ \cline{2-3}
           & 3 & 3 \\
           & 4 & 3 \\
           & 4 & 4 \\
           & 3 & 3 \\ \cline{2-3}
           & 1 & 1 \\
           & 1 & 2 \\
           & 3 & 1 \\
           & 1 & 1 \\ \hline
        \end{tabular}
        \caption{Column Representation}
    \end{subfigure}
    \caption{Column representation for a MultiPolygon}
    \label{fig:geometry-multipolygon}
\end{figure}

A MultiPolygon consists of a sequence of polygons. Each polygon contains one outer shell and zero or more inner holes. In SpatialParquet, we represent a MultiPolygon by storing each ring as a separate part, i.e., exactly in the same way as a regular Polygon. To be able to know where each new polygon starts, we follow the convention of storing outer shells in CW order and inner holes in CCW order. Thus, when reading a MultiPolygon back, after reading each ring, we first test whether the coordinates are stored in CW or CCW order which is a linear-time operation. If it is an inner hole, we append it to a list of rings. If it is an outer shell, we first create a polygon from existing rings and then add the new ring as the first one in the next polygon.

Figure~\ref{fig:geometry-multipolygon} shows an example of a MultiPolygon, $g_6$, with two sub-polygons. There is a total of three rings as shown in the column representation. When parsing this geometry back, SpatialParquet will read the first ring and keep it in a buffer. Then, it will read the second ring and test the order of the points. It will find that they are stored in CCW order so it will keep that ring on the side as a hole. Then, it will read the third ring and find that it is stored in CW order which indicates that it is an outer hole. Thus, it will create the first sub-polygon with one hole. Finally, since no more parts are left in this MultiPolygon, it will create the second sub-polygon with one outer shell and no holes.

\subsection{GeometryCollection}
A GeometryCollection consists of a set of geometries that each can be any of the six geometry types described above as well as another GeometryCollection. Supporting this type is tricky since PBF and Parquet do not allow recursive definition, i.e., a GeometryCollection within another GeometryCollection. We can partially support GeometryCollection by making two changes. First, we change the original definition of Geometry to make the entire Geometry a repeated group. Second, before storing any GeometryCollection we first flatten it by replacing each sub-GeometryCollection with its contents. This way, we remove all recursive definitions in the GeometryCollection and store all sub-geometries at one level. Finally, each geometry in the GeometryCollection is stored as described above.

\section{The Encoding}
\label{sec:encoding}

One of the main advantages of column-oriented stores, is that it groups together homogeneous values, i.e., from the same domain, in each column and makes use of the redundancy among these values to store them more efficiently. This is done through special encoding schemes that work at their best when encoding homogeneous values. For example, a very popular encoding is {\em delta encoding} which stores the deltas (differences) between consecutive values to reduce the storage overhead if the differences are usually small.

This section describes how we encode the column-represented geometries in SpatialParquet to improve the storage efficiency. In SpatialParquet, we primarily have two column types, the geometry {\tt type} column and the {\tt coordinate} columns, which we describe below.

\subsection{Geometry {\tt type} encoding}
Geometry type is an integer value that takes a value in the range $[0, 6]$. In almost all practical cases, all geometries in one dataset have the same type. For example, a point-of-interest dataset will consist of only points. Therefore, we use run-length-encoding (RLE) to encode the geometry type value. RLE replaces consecutive entries with the same value with two numbers, count and value. The former records how many times the latter value is repeated. For example, if all the dataset consists of a single geometry type, e.g., polygons, this column will be stored in SpatialParquet as a pair $(c,3)$, where $c$ is the total number of records in the file and {\tt type=3} is the marker of the polygon data type. Thus, this method can reduce the storage overhead of the {\tt type} column to virtually a constant that does not depend on data size.

\subsection{Geometry {\tt coordinate} encoding}
\label{sec:encoding:coordinate}
The $x$ and $y$ coordinates are stored in floating-point representation\footnote{This paper assumes 64-bit IEEE double floating-point representation but the discussion seamlessly applies to 32-bit single floating-point representation.} A very popular encoding for integer values is delta encoding which stores the first value in a column in full, and then for subsequent values it stores only the delta between each value and its previous one. For example, to store the sequence $\langle 15,16,15,17,20\rangle$, they are replaced with the sequence $\langle 15,+1,-1,+2,+3 \rangle$. The key idea behind delta encoding is that when the deltas have a smaller magnitude value, they can be represented in fewer bits. Additionally, since all the deltas are stored consecutively in the column format, they are bit-packed to reduce the storage size.

Unfortunately, delta encoding can only be directly applied to {\em integer} values. In the IEEE floating point data representation, a smaller magnitude value does not necessarily need fewer bits. This is because any floating point value has to be represented in the (sign, exponent, fraction) format. Also the value has to be first normalized to move the {\em decimal point} right after the most significant one in the number. Check Appendix~\ref{sec:preliminaries} for more details.

When looking at the geometry coordinates, we observe that subsequent values are usually close to each other. For example, a trajectory represented as a MultiPoint is expected to have geographically nearby values. Thus, for both $x$ and $y$ coordinates, each two consecutive values will have a very small difference. However, as mentioned earlier, if we just compute the floating-point difference, we cannot directly reduce the number of significant bits in the number. However, we make another observation that subsequent values are mostly within the same order of magnitude. In other words, they are expected to have either the same, or very close exponents in their floating point representation. Furthermore, if they have the same exponent, then their fractions are also expected to have a small difference.

\paragraph{FP-delta Encoding:}
Based on the observations above we have above, we proposed a floating-point-delta encoding, FP-delta, that requires only one single operation to calculate. FP-delta simply calculates the difference of the integer interpretation of the floating point values. In other words, we ignore the (sign, exponent, fraction) representation and just treat the entire 64-bit double floating-point value is a 64-bit two's complement long integer value. Of course, the difference in this case does not necessarily hold any physical meaning. However, since the exponents are in the most significant part of the value, and if the exponents are similar, then they will cancel each other. Furthermore, if they cancel each other, the resulting delta will represent the difference between the two fractions. Thus, if the two values have the same exponent and their values are close to each other, the FP-delta value is expected to have only a few significant bits which allows us to reduce the amount of storage. As in integer-based delta encoding, we follow our FP-delta encoding with zigzag encoding which maps the deltas of $\langle 0, 1, -1, 2, -2, \ldots\rangle$ to the positive-only value of $\langle 0, 1, 2, 3, 4, \ldots \rangle$. This encoding simply removes the leading ones that are present in negative values in the two's complement representation.

To summarize, the algorithm works as follows. Given a sequence of floating-point numbers, we scan all the values to calculate how many bits we need for the deltas (further explained later). Then, we start producing the output by writing the first value in full. After that, we scan the subsequent values and compute the delta followed by the zigzag encoding, again. At this point, if the delta can be stored in the determined number of bits, we store it directly. Otherwise, if it needs more bits, we store a special reset marker and store the value in full.

\begin{algorithm}[t]
\footnotesize
\begin{algorithmic}[1]
\Function{FP-delta-encode}{double[] $X$, BitOutputStream out}
\State $n^*$ = {\sc computeBestDeltaBits}($X$) \label{alg:fp-encode:deltaBits}
\State resetMarker = -1 $\ggg$ (64-$n^*$)
\State significantOnes = -1 $\ll$ ($n^*$)
\State out.write($n^*$, 8)
\State out.write($X[0]$, 64) \label{alg:fp-encode:value0}
\For{$i$ = 1 {\bf to} $|X|-1$} \label{alg:fp-encode:forloop1}
  \State delta = cast-long($X[i]$)-cast-long($X[i-1]$)
  \State zigzag = (delta $\gg$ 63) $\oplus$ (delta $\ll$ 1)
  \If{(zigzag $\&$ significantOnes $\neq 0 $) or (zigzag $=$ resetMarker) }
    \State out.write(resetMarker, $n^*$)
    \State out.write($X[i]$, 64)
  \Else
    \State out.write(zigzag, $n^*$)
  \EndIf
\EndFor
\EndFunction
\end{algorithmic}
\caption{FP-delta encoding algorithm*}
\label{alg:fp-encode}
\begin{flushleft}
* $\gg$ is the arithmetic shift right, $\ggg$ is the logical shift right, $\ll$ is shift left, $\&$ is the logical AND operator, and $\oplus$ is bit-wise XOR
\end{flushleft}
\end{algorithm}

Algorithm~\ref{alg:fp-encode} provides the pseudo-code of the FP-delta encoding algorithm. The input is an array of floating-point values and the output goes to a bit output stream. The bit output stream bit-packs all the written values depending on how many bits are used for each value. Line~\ref{alg:fp-encode:deltaBits} uses the function {\sc computeBestDeltaBits}, explained later, to compute the number of bits, $n^*$, to use for deltas that will minimize the overall output size. After that, it computes the reset marker which is simply the highest possible zigzag-delta, i.e., all bits set to one. It then computes a value with its most significant bits set to one, corresponding to the number of bits to be discarded. Line~\ref{alg:fp-encode:value0} writes the first value as a full 64-bit value. Then, the loop in Line~\ref{alg:fp-encode:forloop1} writes all the remaining values. First, it computes the FP-delta and encodes it using zigzag. Then, it checks if the zigzag-encoded delta has any of its significant bits set to one out of the bits to be removed or if it is equal to the reset marker. If so, we first write the reset marker followed by the full 64-bit value. Otherwise, it writes this value directly to the output stream in $n^*$ bits.

To complete the encoding algorithm, we describe how to compute the number of bits that minimizes the storage size. The challenge in this part is to balance the trade-off between the number of bits needed for deltas and the number of overflow deltas that need to be encoded as 64-bit values. To overcome this challenge, we scan the entire dataset to compute the number of bits needed for each delta. As we do so, we build a histogram $h$ of number of deltas for each number of bits. That is $h[n]$ indicates the number of deltas that need at least $n$ bits to be stored. This means that $\sum{h}=|X|-1$ since each value falls into one of the histogram bins.
Then, we calculate the output size for each of the deltas and choose the minimum.

To calculate the output size $S$, recall that each value that can be encoded as a delta will take $n$ bits, where $n$ is the number of bits reserved for the deltas. Any value that cannot be encoded in $n$ bits will be stored as $n+64$ bits, that is, the special marker followed by the full 64-bit value. Thus, for any value $n$, the output size can be calculated using the following equation.

\begin{equation}
S(n)=n\cdot\sum_{i=0}^{i=n}{h[i]}+(n+64)\cdot\sum_{i=n+1}^{i=64}{h[i]}
\end{equation}

Since $\sum{h}=|X|-1$, we can rewrite this equation as follows.

\begin{equation}
S(n)=n\cdot (|X|-1)+64\cdot\sum_{i=n+1}^{i=64}{h[i]}
\label{eqn:outputsize}
\end{equation}

The optimal number of bits for the deltas to minimize the storage size is as follows.

\begin{equation}
n^*=\argmin_{0\le n \le 64}{\{S(n)\}}
\end{equation}

There are two notes about this approach. First, since we can accurately compute the encoded data size, we can easily skip this algorithm altogether and store the data in its raw format in case the calculated saving is very little. Second, this algorithm has space and time complexities of $O(|X|)$ since it needs to store and scan all the values before writing them. However, keep in mind that this algorithm is applied for each {\em page} in the Parquet file which has a default size of 1MB. Thus, the overhead is not significant. Still, if we want to save the overhead, we can skip this function and use an empirical best value that can be calculated experimentally for the application depending on the dataset characteristics. The pseudo-code for this algorithm is available in Appendix~\ref{sec:pseudo-code}.

\paragraph{FP-delta Decoding:}
The decoding algorithm is simpler since it does not need to calculate the output size or choose the best number of bits for the delta. The decoder begins by reading the first byte that contains the number of bits $n$ that will be used for the deltas. Then, it reads a 64-bit value that represents the first-floating point value. After that, it keeps reading $n$ bits from the input and checking the value. If it is not the reset marker, it is treated as delta and the next value is emitted accordingly. Otherwise, if it is the reset marker, it is ignored and the next 64-bit double value is read and emitted.

\begin{algorithm}[t]
\footnotesize
\begin{algorithmic}[1]
\Function{FP-decode}{BitInputStream in, DoubleOutput out}
\State $n$ = in.read(8)
\State reset-marker = -1 $\gg$ (64-$n^*$)
\State prev-x = cast-double(in.read(64))
\State out.write(x)
\While{more values in the input}
  \State zigzag = in.read($n$)
  \If{zigzag $\neq$ reset-marker}
    \State delta = (zigzag $\ggg$ 1) $\oplus$ -(zigzag $\&$ 1)
    \State next-x = cast-double((cast-long)prev-x + delta)
  \Else
    \State next-x = cast-double(in.read(64))
  \EndIf
  \State out.write(next-x)
  \State prev-x = next-x
\EndWhile

\EndFunction
\end{algorithmic}
\caption{FP-delta decoding algorithm}
\begin{flushleft}
$\ggg$ is logical shift right, $\&$ is bit-wise AND
\end{flushleft}
\label{alg:fp-decode}
\end{algorithm}

Algorithm~\ref{alg:fp-decode} gives the pseudo-code of the FP-delta decoding algorithm. It takes an input bit stream and produces all the output values to a DoubleOutput stream. First, it reads an 8-bit value that indicates the number of bits $n$ used for the deltas. Then, it reads the first value as a 64-bit floating point value and emits it directly to the output. It also keeps it in the variable $prev-x$ to use for delta decoding. After that, it keeps reading from the input bit stream as long as there are value. Notice that Parquet determines if there are more values depending on the definition level so we do not need to keep this information. For each value, it first reads $n$ bits that represent the zigzag-encoded delta. If this value is not equal to the reset marker, it reverses the zigzag encoding to get the delta value as shown. The next value is then calculated by adding this delta to the previous value. Notice that the addition is a 64-bit integer addition and we cast the values accordingly to get the actual double value. Otherwise, if the zigzag value is equal to the reset marker, we simply discard it and use the next 64-bit value as the next $x$ value. Finally, we emit this value and keep it in the buffer so that we can read the next value.

\paragraph{Parquet Integration}
The only encoder for floating point values in Parquet is the raw-encoder that stores each value as-is, and may apply a run-length-encoding to save space on repeated values. To integrate our FP-delta method into Parquet, we extended the {\em Encoder} and {\em Decoder} interfaces in Parquet to support the encoder and decoder functions described above. We configured Parquet to automatically use the FP-delta method for floating-point values in the geometry data type. This allows Parquet to still recognize the floating-point values in the coordinate columns which is helpful for the next part, the indexing.

\section{The Indexing}
\label{sec:indexing}

\begin{figure*}
    \begin{subfigure}{0.48\columnwidth}
        \includegraphics[width=0.9\linewidth]{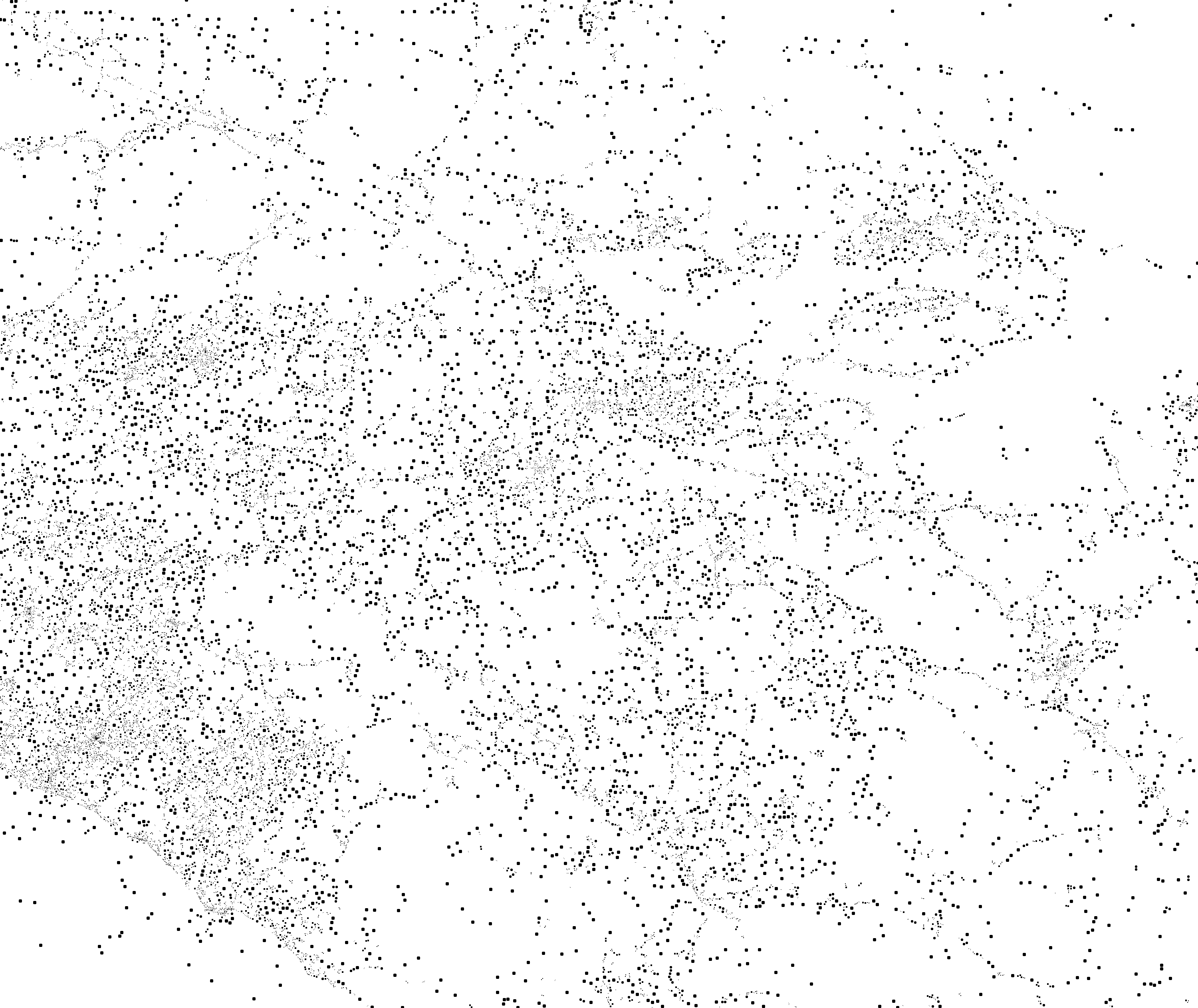}
        \caption{Input data (5.6 M points)}
    \end{subfigure}
    \hfill
    \begin{subfigure}{0.48\columnwidth}
        \includegraphics[width=0.9\linewidth]{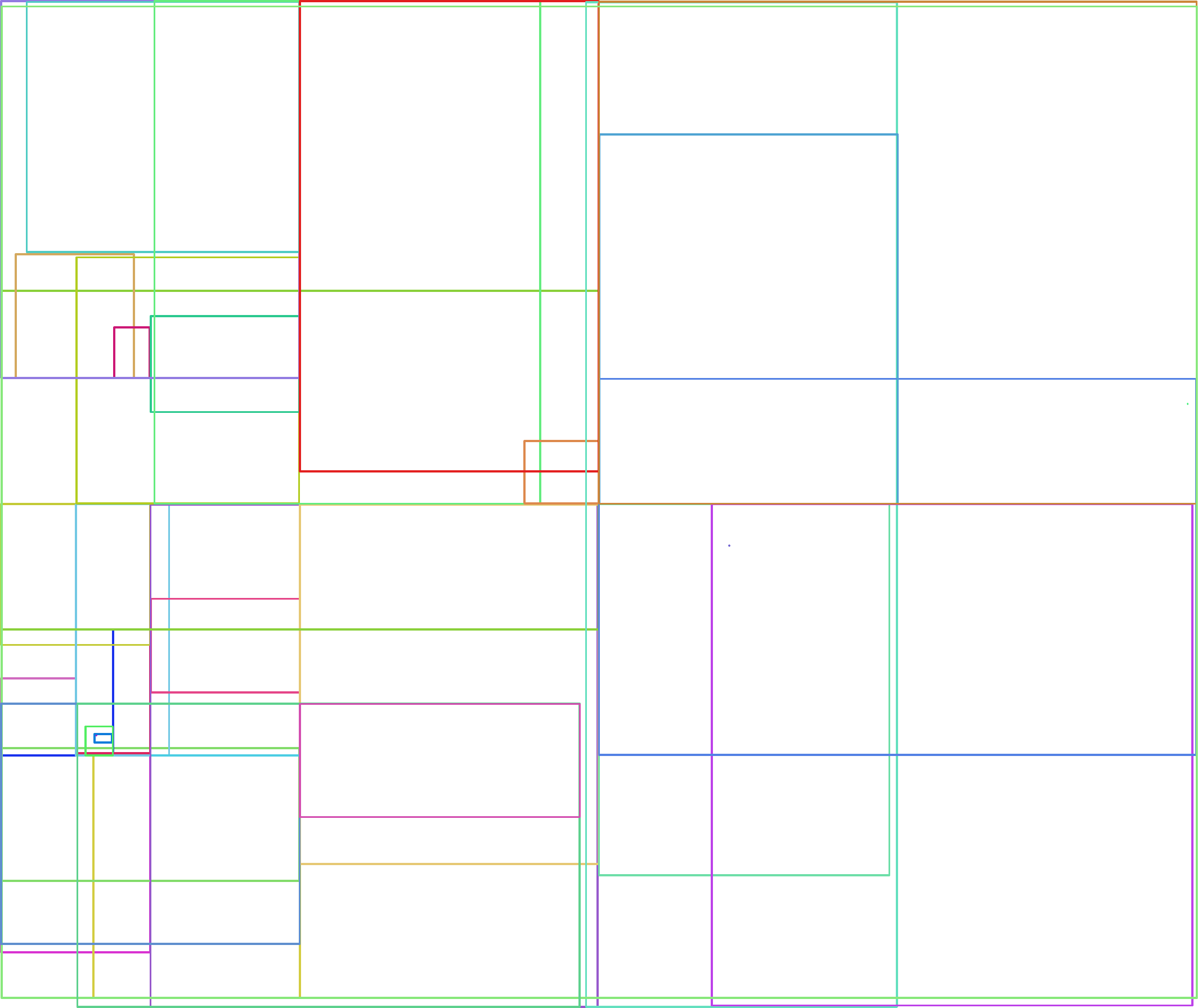}
        \caption{Z-Curve (0.56 seconds)}
    \end{subfigure}
    \hfill
    \begin{subfigure}{0.48\columnwidth}
        \includegraphics[width=0.9\linewidth]{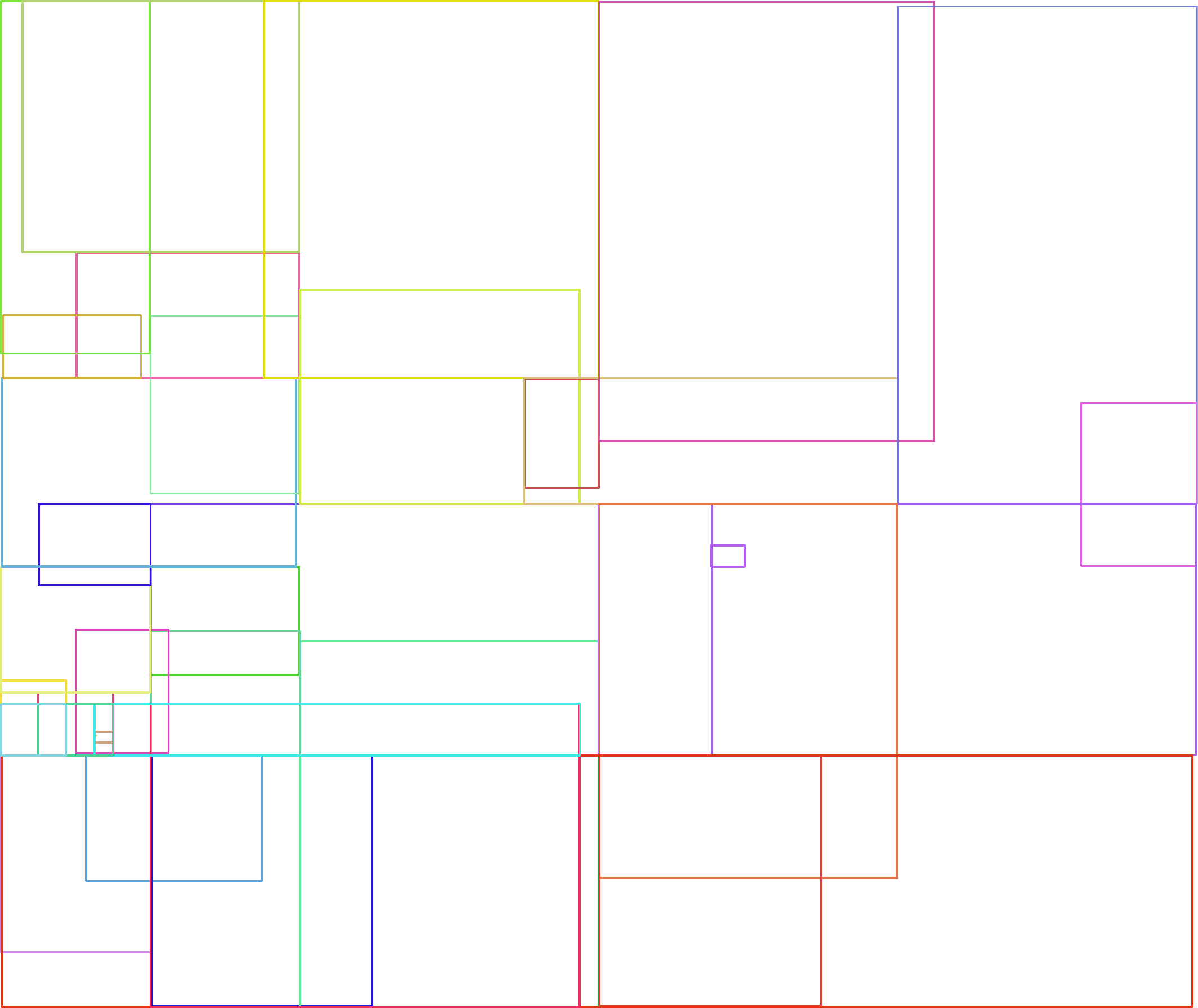}
        \caption{Hilbert-Curve (0.74 seconds)}
    \end{subfigure}
    \hfill
    \begin{subfigure}{0.48\columnwidth}
        \includegraphics[width=0.9\linewidth]{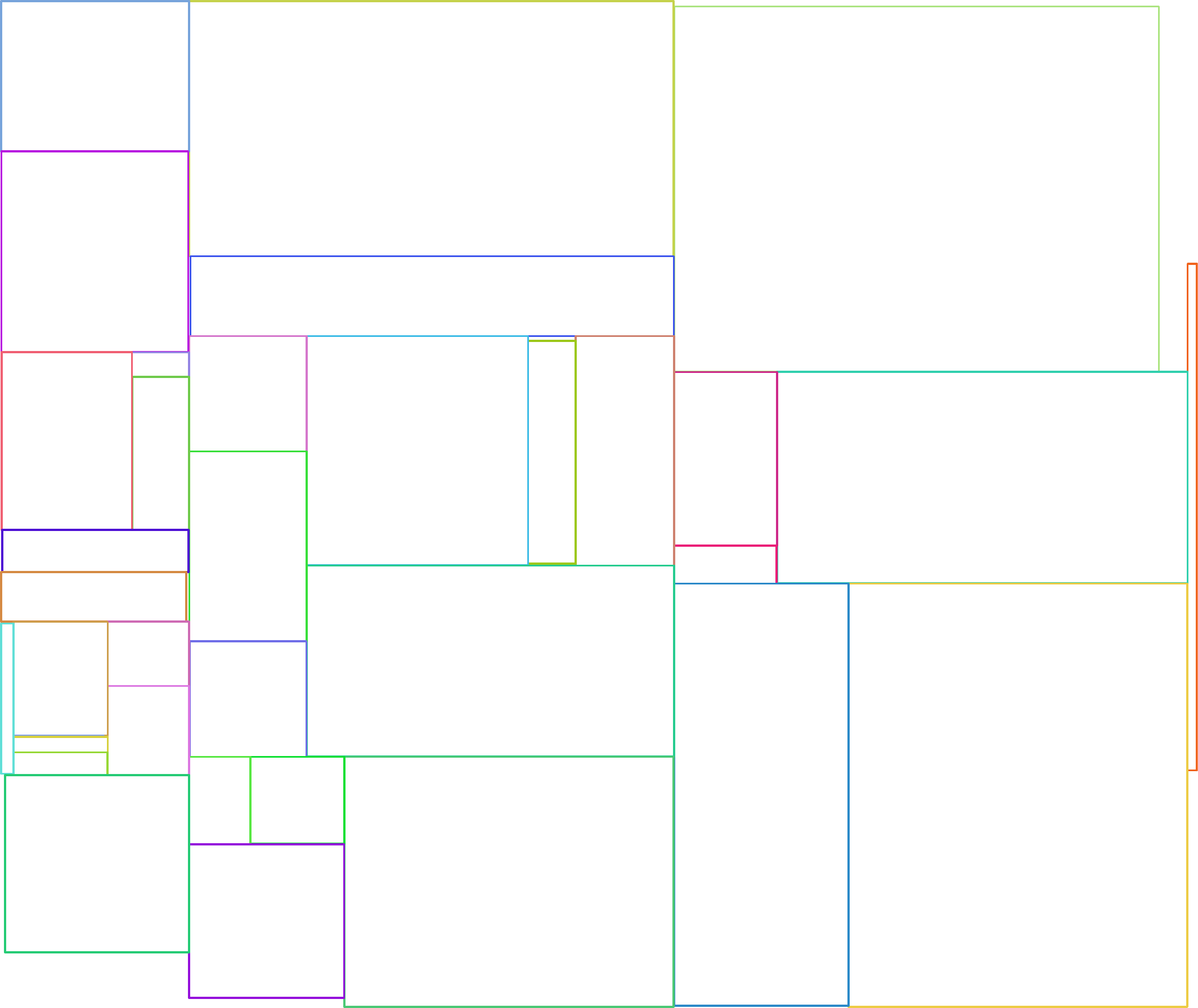}
        \caption{R*-Grove (9.47 seconds)}
    \end{subfigure}
    \caption{The effect of sorting on page boundaries with three methods with their respective running time}
    \label{fig:partitions}
\end{figure*}

This section describes how to build a light-weight index on-top of SpatialParquet using the built-in structure of Parquet. The goal of the index is to be able to avoid reading the entire file if the user wants to process data in a small geographical region. Most popular spatial indexes, e.g., Quad-tree, R-tree, and GridFile, group records into minimum bounding rectangles (MBR) so that an entire block can be skipped when it falls outside the query range.

Parquet provides a light-weight method of pruning non-relevant records by adding {\em column statistics}. Mainly, it adds the range, {\em [min, max]}, for each column and allows the user to skip reading the data if the desired range does not overlap with the column data range. To make this pruning more effective, Parquet splits each column into pages of roughly 1MB each so that it can skip over some of these pages. Furthermore, each page is compressed separately which makes skipping these pages more efficient.

The structure we propose in SpatialParquet gives us the opportunity of collecting statistics for the $x$ and $y$ columns. This is only possible because Parquet identifies each of these as a separate column. In contrast, if we store the entire geometry in WKB or WKT format, Parquet will not be able to collect these statistics. Together, the ranges of $x$ and $y$ make a spatial bounding box for each page. Thus index construction is as simple as instructing Parquet to collect the minimum and maximum for the $x$ and $y$ columns and store them in the output file.

At reading time, if the user provides a query rectangle in the form of $[(x_{min},y_{min}),(x_{max}, y_{max})]$, we translate it into two separate ranges, $[x_{min},x_{max}]$ and $[y_{min},y_{max}]$ for the $x$ and $y$ columns, respectively. Then, we pass these ranges to Parquet to read only the pages that intersect the query range.

The effectiveness of this light-weight index highly depends on the distribution of the data. in one extreme, if data is highly scattered such that each page contains data from all over the data domain, then this technique will be very ineffective since all $[min,max]$ ranges will cover almost the entire input space leaving no room for pruning. On the other extreme, this technique will be very efficient if the data is highly-clustered, that is, spatially nearby records are very close to each other in the input order. This will allow the range of each page to be very tight and more pruning can happen.

To improve the effectiveness of the index, we add a sorting step that tries to cluster nearby records. Notice that it does not have to be perfect. All we need is to avoid the very bad situation where each page covers the entire world. Thus, we do not want to pay a high cost for a very accurate partitioning technique. We decided to use two light-weight space filling curve sort methods, namely, Z-curve and Hilbert curve. Both techniques can provide a linear sort order that takes into account both $x$ and $y$ coordinates, with Hilbert curve known to be more effective with a slightly higher computation cost. Furthermore, since this is not a traditional hierarchical index with a single root, we do not care about sorting the entire dataset which can be very costly. Rather, we process the records into groups with a fixed number of records, e.g., one million. Whenever we have that number of records, we sort them and write them to SpatialParquet. This ensures that the memory overhead is upper bounded and that the computation overhead of sorting grows only linearly with the data size.

Figure~\ref{fig:partitions} provides an example of the effect of sorting the records before writing the SpatialParquet file for a 5.6 million point dataset. Without using any sorting, all partitions cover the entire input space so we do not show it here. While not perfect, both of them provide an opportunity for pruning some disk pages depending on the range query. Notice that SpatialParquet does not care how the partitions are created so if the user is willing to spend more time, more sophisticated big-data partitioning techniques can be used~\cite{EAM15}. In the figure, we show an example of using R*-Grove~\cite{VE20b} which yields much better partitions with no overlap but, in this case, it takes slightly less than 10 seconds. Given the static nature of the data, users might also be interested in using learned spatial indexes~\cite{AWA20} but we leave this as future research directions.

\section{Experiments}
\label{sec:experiments}

This section shows the results of an extensive experimental evaluation that compares SpatialParquet to existing spatial data formats, as well as, studying the effect of various parameters.

All experiments are executed on a machine running on Intel(R) Xeon(R) CPU E5-2603 v4 @ 1.70GHz and 64GB of RAM.

We use four datasets for all evaluations. All the dataset are publicly available on UCR-Star~\cite{GVE+19} and are summarized in Table~\ref{tab:datasets}. Each of these datasets contains predominantly one geometry type. This helps us evaluate the effect of different geometry types on compression and storage size. These datasets also have different sizes in terms of the total number of geometries and points contained in them. The versions of these datasets only contain geometry data, and all objects are stripped of any metadata. This so that we compare purely on the geometry values, since this is the main focus of our work. Second, the objects on some of these datasets already follow some sorted order as they are provided from the source. We discuss the effect of sorting in detail. 

We implement SpatialParquet in Java based on the original Parquet Java repository~\cite{noauthor_parquet_2022}.

\begin{table}[t]
\centering
\caption{Experiment Datasets}
\label{tab:datasets}
\resizebox{\columnwidth}{!}{%
\begin{tabular}{l|llll}
Dataset (Acronym)  & Geometry Type   & \# of Geometries & Num points \\ \hline
Porto Taxi (PT)    & MultiPoint      &  1.7 M           &  83 M      \\
TIGER18/Roads (TR) & MultiLineString &  18 M            &  350 M     \\
MSBuildings (MB)   & Polygon         &  125 M           &  753 M     \\
eBird (eB)        & Point           &  801 M           &  801 M   \\ \hline           
\end{tabular}%
}
\end{table}

In the remainder of this section, first we compare our proposed work to existing spatial formats. We discuss both savings in terms of required storage size, as well as, performance implications. After that, we show the effect of space-filling-curve-based sorting on the distribution of values in each column chunk. Following, we show the effect of various parameters like encoding, compression, and sorting on the data size and on the performance. Finally, we show how column statistics improves the performance of reading by pruning column chunks and pages based on a provided filtering range.

\subsection{Comparing to Existing Spatial Formats}

We compare Spatial Parquet to three existing baselines: GeoParquet, ShapeFile, and GeoJSON. We use Java implementations for all of these baselines. For GeoParquet, its existing implementation is only available as a Python package, so we provide a new Java implementation for a fair comparison. Its implementation differs from SpatialParquet in which it requires five values per geometry object. One value represents the Well-Known-Binary (WKB) of the geometry, and the other four values determine the minimum-bounding-rectangle of the geometry for easy filtering. For reading and writing ShapeFile and GeoJSON we use implementations provided in Beast~\cite{eldawy2021beast}.

For evaluations in this section, the source data for writing these files are sorted using the Hilbert-curve method. The details of how the sorting is applied and its effects are provided in the following sub-section.

In the first evaluation, we evaluate the formats based on the total data size without any compression. The left part of Table~\ref{tab:baselines_datasize} shows the size of data stored in these formats without any compression. For all datasets, SpatialParquet with delta encoding significantly decreases the size of the data. In the case of the PT, MB and eBird datasets, its size is less than half that of the nearest data size for the other formats. We believe that this alone makes SpatialParquet a strong candidate for storing Geospatial data.

Compressing the data using a general purpose compression techniques can further reduce the sizes, but it has some performance implications for both reading and writing. We store the same files in the previous evaluation after compressing their contents using GZIP~\cite{deutsch1996gzip} compression technique. However, there are some differences in how the compression is applied. In SpatialParquet and GeoParquet, the compression is applied on column pages, usually one megabyte in size. This is important to minimize the overhead when reading records from an arbitrary position. Also, compressing small chunks of data adds minimal overhead at write time, but it loses a little bit compared to when the compression is applied to the entire dataset at once. For GeoJSON files, the entire dataset is written as one giant {\tt .geojson.gz} file. Thus, the compression technique is applied to it as one large file. ShapeFiles are compressed a little differently. Due to the implementation used for writing Shapefiles, to avoid running out of memory, we divide the data up to one million geometry object and write them into a separate Shapefile partition. So when we apply the compression, we apply it to each of these files individually. The result of this evaluation is shown in the right part of Table~\ref{tab:baselines_datasize}. Clearly, compression adds significant benefits for all formats, with SpatialParquet still considerably lower than all other formats, except in two cases. The compressed GeoJSON file of the MSBuildings dataset is slightly smaller than that of GeoParquet. This can be attributed to the fact that compression is applied differently as described. This shows that compressing an entire dataset using GZIP can save a little bit in storage size compared to compressing small chunks separately. However, keep in mind that a compresses GeoJSON file has to be processed entirely and sequentially while SpatialParquet provides more efficient access methods such as choosing specific column or filtering by rows using the light-weight index (as detailed shortly). Other binary formats are up-to three times bigger than SpatialParquet.
\begin{table}[]
\centering
\caption{Output size in GB with/without compression}
\label{tab:baselines_datasize}
\resizebox{\columnwidth}{!}{%
\begin{tabular}{|l|llll||llll|}
\hline
\textbf{} &
  \multicolumn{4}{l||}{\textbf{Uncompressed}} &
  \multicolumn{4}{l|}{\textbf{Compressed}} \\ \hline
\textbf{Format} &
  \multicolumn{1}{l|}{\textbf{PT}} &
  \multicolumn{1}{l|}{\textbf{TR}} &
  \multicolumn{1}{l|}{\textbf{MB}} &
  \textbf{eB} &
  \multicolumn{1}{l|}{\textbf{PT}} &
  \multicolumn{1}{l|}{\textbf{TR}} &
  \multicolumn{1}{l|}{\textbf{MB}} &
  \textbf{eB} \\ \hline
SpatialParquet &
  \multicolumn{1}{l|}{\textbf{0.856}} &
  \multicolumn{1}{l|}{\textbf{3.5}} &
  \multicolumn{1}{l|}{\textbf{8.2}} &
  \textbf{11} &
  \multicolumn{1}{l|}{\textbf{0.388}} &
  \multicolumn{1}{l|}{\textbf{1.9}} &
  \multicolumn{1}{l|}{4.0} &
  1.9 \\ \hline
GeoParquet &
  \multicolumn{1}{l|}{1.8} &
  \multicolumn{1}{l|}{6} &
  \multicolumn{1}{l|}{17} &
  43 &
  \multicolumn{1}{l|}{0.718} &
  \multicolumn{1}{l|}{3.5} &
  \multicolumn{1}{l|}{8.7} &
  6 \\ \hline
ShapeFile &
  \multicolumn{1}{l|}{1.4} &
  \multicolumn{1}{l|}{6.4} &
  \multicolumn{1}{l|}{19} &
  28 &
  \multicolumn{1}{l|}{0.654} &
  \multicolumn{1}{l|}{3.5} &
  \multicolumn{1}{l|}{7.8} &
  5.7 \\ \hline
GeoJSON &
  \multicolumn{1}{l|}{2.2} &
  \multicolumn{1}{l|}{14} &
  \multicolumn{1}{l|}{32} &
  97 &
  \multicolumn{1}{l|}{0.439} &
  \multicolumn{1}{l|}{2.2} &
  \multicolumn{1}{l|}{\textbf{3.8}} &
  \textbf{1.8} \\ \hline
\end{tabular}%
}
\end{table}

Next, we compare the writing time of SpatialParquet against the baselines. Table~\ref{tab:baselines_time} shows the writing time in seconds for the uncompressed files. SpatialParquet has the best performance by far for the PT and eB datasets. However, it performs slower than GeoParquet on the TR and MB datasets. These two dataset contain geometries of type MultiLineString and Polygon, respectively. These two data-types are more complex than the Point and MultiPoint types. More complex types require more calls to the parquet interface, since we send each individual value by itself, and it has to track the size of each geometry part. Because Parquet has BYTE ARRAY as a native type the well-known-binary (WKB) is sent directly as one value through the Parquet interface. Keep in mind that our current implementation is a first-cut solution while WKB reading and writing has been optimized for years. Given the huge space saving of SpatialParquet, we will further optimize the writing operation to reduce any potential overhead.

Finally, we compare SpatialParquet to the baselines in terms of the reading time. Table~\ref{tab:baselines_time} shows the reading time in seconds for the uncompressed files. GeoParquet and Shapefile have the best reading times. Similar to writing, reading data in WKB is much more efficient than requesting values repeatedly through the Parquet interface. We believe we can improve the reading performance in the future by providing a lower-level access to the coordinate arrays from Parquet rather than reading one value at a time using the current API. Notice that the information is already stored in the Parquet file in the format that we want, i.e., consecutive arrays of floating points values, so we should be able to further optimize the reading part in the future without changing the SpatialParquet format.

\begin{table}[]
\centering
\caption{Write/Read time in seconds for uncompressed formats}
\label{tab:baselines_time}
\resizebox{\columnwidth}{!}{%
\begin{tabular}{|l|llll||llll|}
\hline
\textbf{} &
  \multicolumn{4}{l||}{\textbf{Writing Time}} &
  \multicolumn{4}{l|}{\textbf{Reading Time}} \\ \hline
\textbf{Format} &
  \multicolumn{1}{l|}{\textbf{PT}} &
  \multicolumn{1}{l|}{\textbf{TR}} &
  \multicolumn{1}{l|}{\textbf{MB}} &
  \textbf{eB} &
  \multicolumn{1}{l|}{\textbf{PT}} &
  \multicolumn{1}{l|}{\textbf{TR}} &
  \multicolumn{1}{l|}{\textbf{MB}} &
  \textbf{eB} \\ \hline
SpatialParquet &
  \multicolumn{1}{l|}{\textbf{74}} &
  \multicolumn{1}{l|}{215} &
  \multicolumn{1}{l|}{544} &
  \textbf{833} &
  \multicolumn{1}{l|}{49} &
  \multicolumn{1}{l|}{143} &
  \multicolumn{1}{l|}{455} &
  546 \\ \hline
GeoParquet &
  \multicolumn{1}{l|}{226} &
  \multicolumn{1}{l|}{\textbf{99}} &
  \multicolumn{1}{l|}{\textbf{425}} &
  1490 &
  \multicolumn{1}{l|}{\textbf{17}} &
  \multicolumn{1}{l|}{64} &
  \multicolumn{1}{l|}{204} &
  \textbf{500} \\ \hline
ShapeFile &
  \multicolumn{1}{l|}{123} &
  \multicolumn{1}{l|}{490} &
  \multicolumn{1}{l|}{1445} &
  4246 &
  \multicolumn{1}{l|}{88} &
  \multicolumn{1}{l|}{\textbf{43}} &
  \multicolumn{1}{l|}{\textbf{161}} &
  534 \\ \hline
GeoJSON &
  \multicolumn{1}{l|}{105} &
  \multicolumn{1}{l|}{485} &
  \multicolumn{1}{l|}{956} &
  2342 &
  \multicolumn{1}{l|}{55} &
  \multicolumn{1}{l|}{424} &
  \multicolumn{1}{l|}{610} &
  1280 \\ \hline
\end{tabular}%
}
\end{table}

In summary, SpatialParquet provide an excellent alternative for the existing geo-spatial data formats. It can add significant savings in storage requirements. It does not currently always perform the best in terms of writing and reading speed for all datasets, but future improvements can be added to bridge this gap without changing the proposed format. In addition to adding more improvements to the implementation, we plan to integrate it within Beast~\cite{eldawy2021beast}, which would make it more straightforward to convert from and to SpatialParquet with other geo-spatial formats, on top of making it possible to process big data in distributed systems.

\subsection{Effect of Sorting on Sample Distribution}

\begin{figure}[t]
    \centering
    \begin{subfigure}{0.49\linewidth}
        \includegraphics[width=\linewidth]{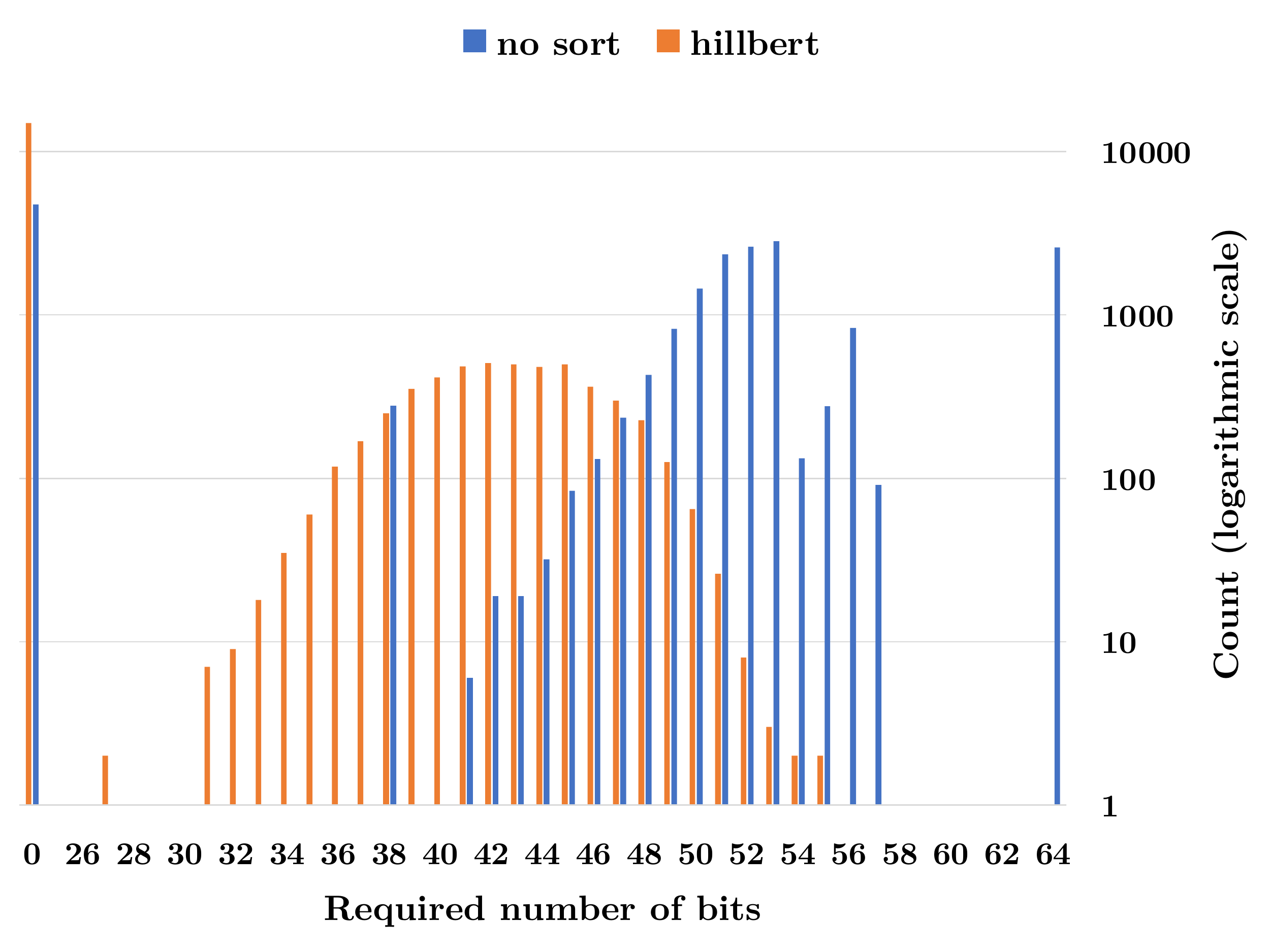}
        \caption{eBird}
        \label{fig:eBird_dist}        
    \end{subfigure}
    \begin{subfigure}{0.49\linewidth}
        \includegraphics[width=\linewidth]{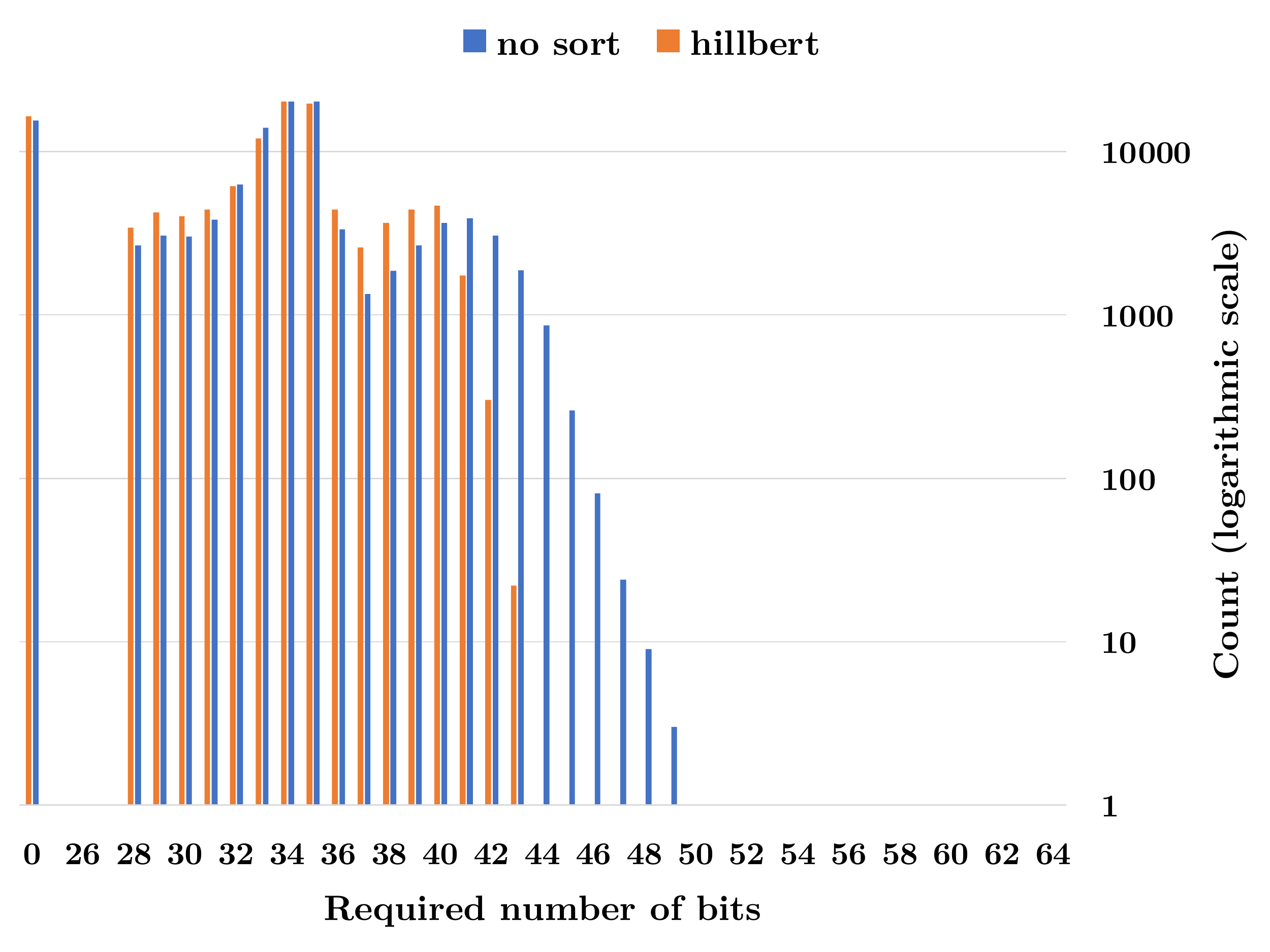}
        \caption{MSBuildings}
        \label{fig:MSBuildings_dist}
    \end{subfigure}
    \caption{The effect of sorting on the number of records that require at least $n$ bits for delta encoding}
    \label{fig:delta-distribution}
\end{figure}

In this section, we highlight the effect of sorting on the distribution of the deltas in a column chunk. Sorting results in delta values with less number of significant bits. We show the histogram of the first column chunk of the eBird dataset and the MSBuildings dataset. Sorting has a considerable effect on the eBird dataset because it is not readily sorted from the source. The MSBuildings dataset is somewhat sorted because the data is divided by U.S. state, but applying a sort function still adds more benefit. The other two datsets seem to be well sorted from the source and do not benefit much from sorting. Figure~\ref{fig:eBird_dist} shows how a Hilbert-curve sort shifts the distribution of values towards the left, meaning more values requiring less number of significant bits. The count of delta values that require 64 significant bits is large for the unsorted eBird dataset because it includes consecutive values alternating between positive and negative numbers. However, this basic sorting completely removes this spike. We notice the same effect on the MSBuildings dataset in Figure~\ref{fig:MSBuildings_dist}, but to a lesser degree since all records are in the US. Applying a Z-curve sort, has more or less of a similar effect on all datasets.

Another thing we notice from these two figures is that a considerable number of values are concentrated at the zero significant bit bar. This bar represents consecutive values that are exactly the same. This is common for geotagged values that are created from the same address. This means that we can add an additional run-length-encoding after the deltas to store all of these consecutive equal delta values in a more compact way. We leave this as a future improvement.

\subsection{Evaluating Possible Configurations in SpatialParquet}

\begin{figure}[t]
    \begin{subfigure}{0.49\linewidth}
        \includegraphics[width=\linewidth]{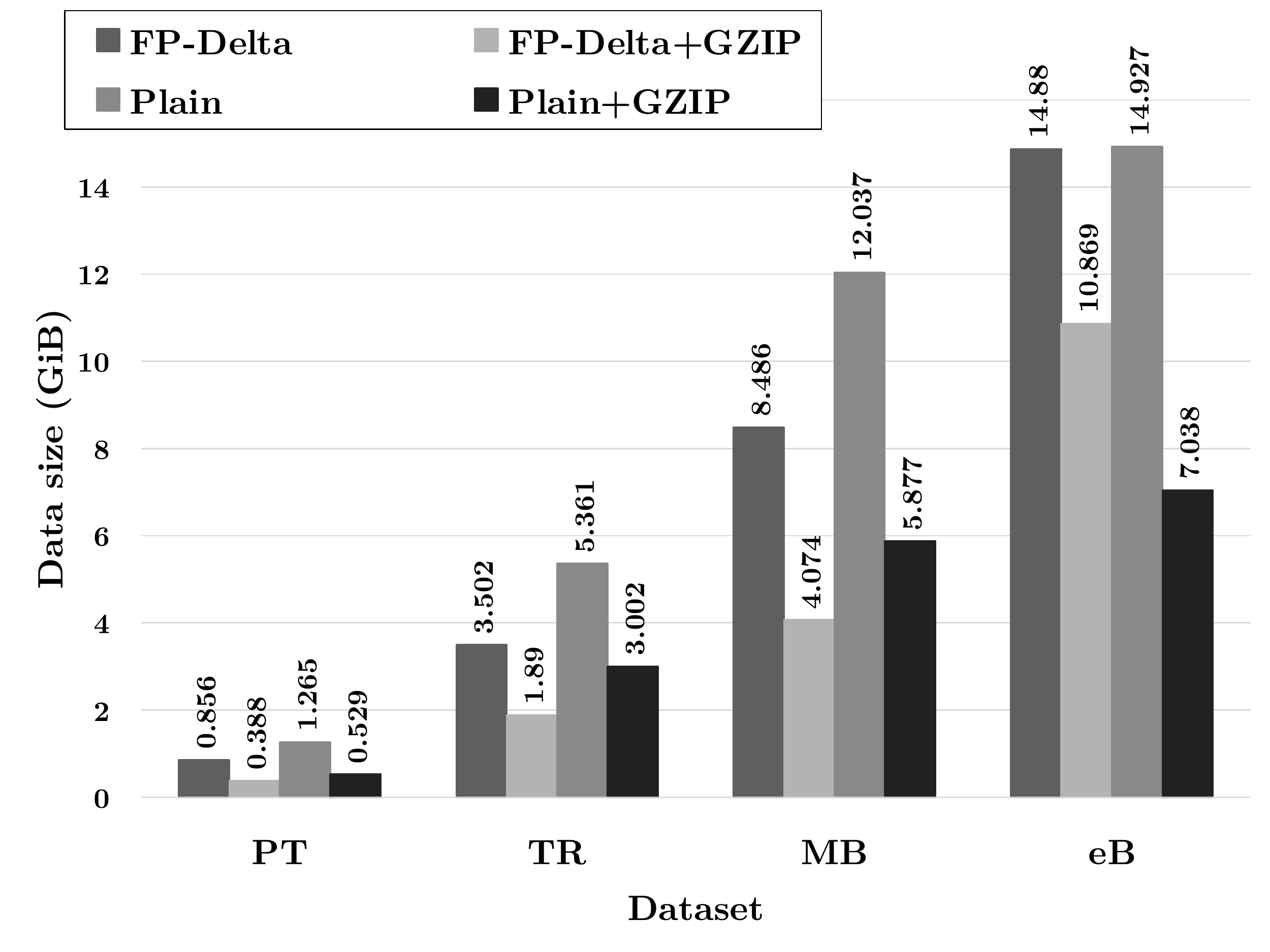}
        \caption{No sorting}
        \label{fig:delta_plain}
    \end{subfigure}
    \begin{subfigure}{0.49\linewidth}
        \includegraphics[width=\linewidth]{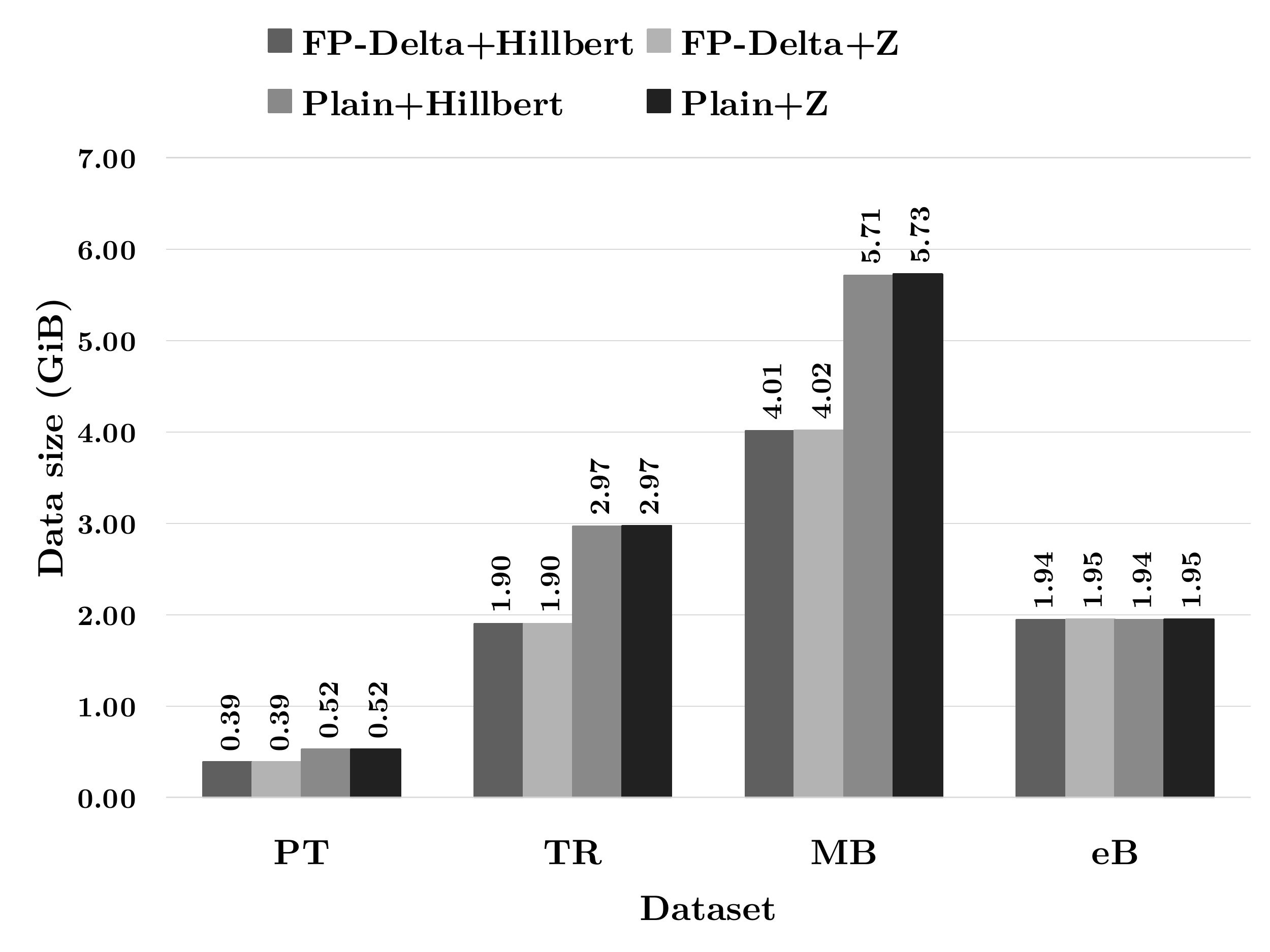}
        \caption{Hilbert- and Z-Curve sorting}
        \label{fig:delta_plain_sort}
    \end{subfigure}
    \caption{The effect of sorting on output size in SpatialParquet}
    \label{fig:sorting-size}
\end{figure}

In this part, we delve into SpatialParquet to evaluate the possible configurations for it.
First, we look into the effect of using FP-delta with and without compression, before applying any sorting. In Figure~\ref{fig:delta_plain}, in all cases FP-delta results in a smaller size with and without GZIP compression, except for the eBird dataset. The eBird is not sorted by default and since all of its geometries are points, there is no gain from applying the delta to a single geometry object. Therefore, sorting is required to significantly reduce the size.

We show the sizes of compressed data after sorting in Figure~\ref{fig:delta_plain_sort}. The main difference can be noticed in the eBird dataset because it is the only one that is not originally sorted, although we could have shuffled the other datasets for the purposes of this experiment.

\begin{figure}[h]
    \centering
    \includegraphics[width=\columnwidth]{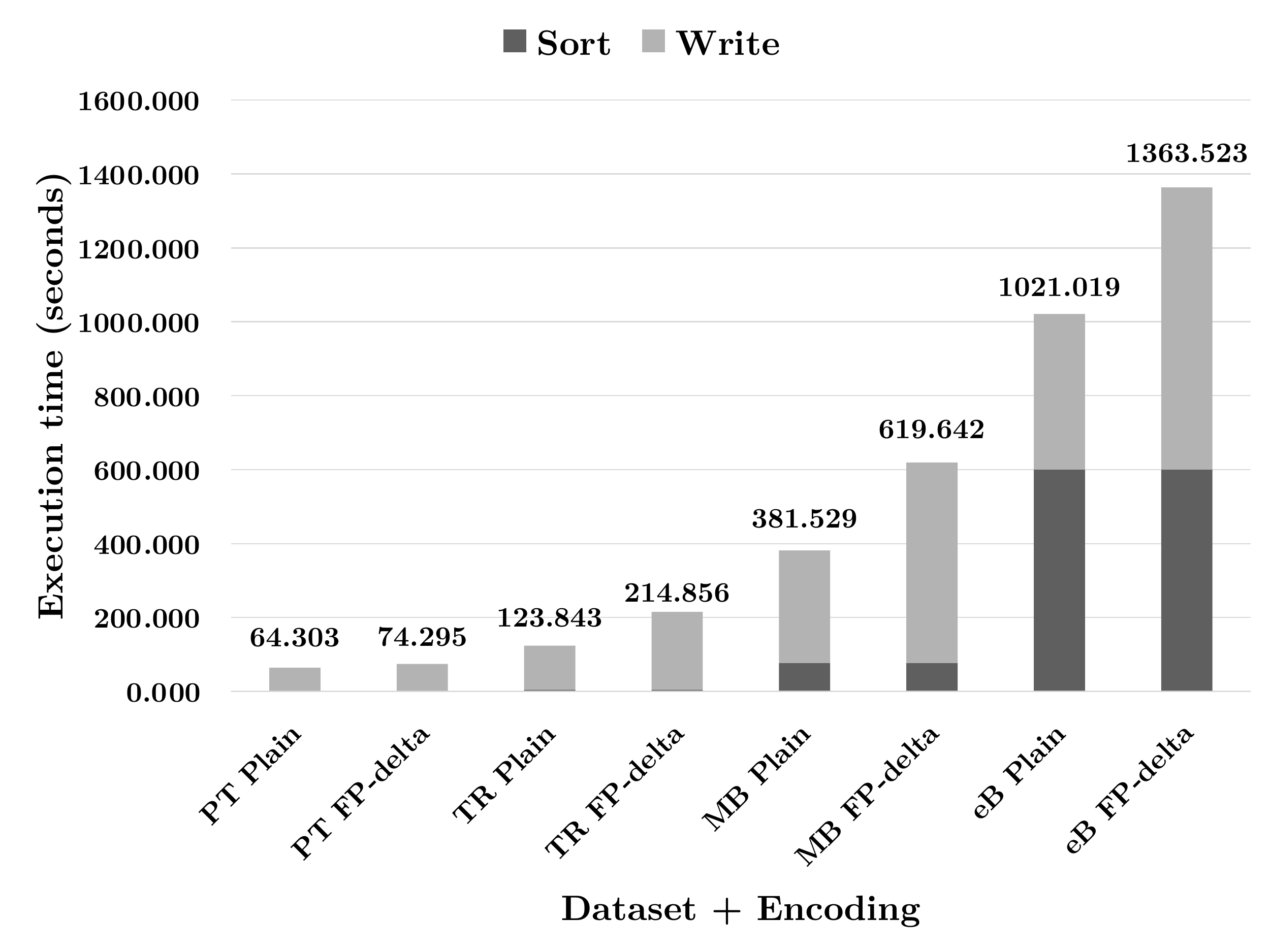}
    \caption{Encoding and sorting overhead}
    \label{fig:write_overhead}
\end{figure}

Both FP-delta and sorting add benefits in reducing the final data size, but they add some performance overhead. This evaluation is depicted in Figure~\ref{fig:write_overhead}. FP-delta requires an additional iteration over the data, given that it calculates the final bit size after completing all the differences. Also, its current implementation involves allocating an additional buffer to store the final values. In the worst case, it seems that it adds up to 80\% of overhead compared to writing the plain double values. Sorting can add a significant overhead, because it is performed sequentially on a buffer of size at most one million objects. However, considering the major benefits it adds this should be negligible. Moreover, in practice we can sort very big data using distributed sorting/indexing techniques. Beast~\cite{eldawy2021beast} has several of these methods implemented on top of Spark. We plan to integrate Spatial Parquet within Beast, which would make sorting/indexing, among other optimizations, a more seamless process.

\subsection{Column Statistics and Filtering}

\begin{figure}[h]
    \centering
    \includegraphics[width=\columnwidth]{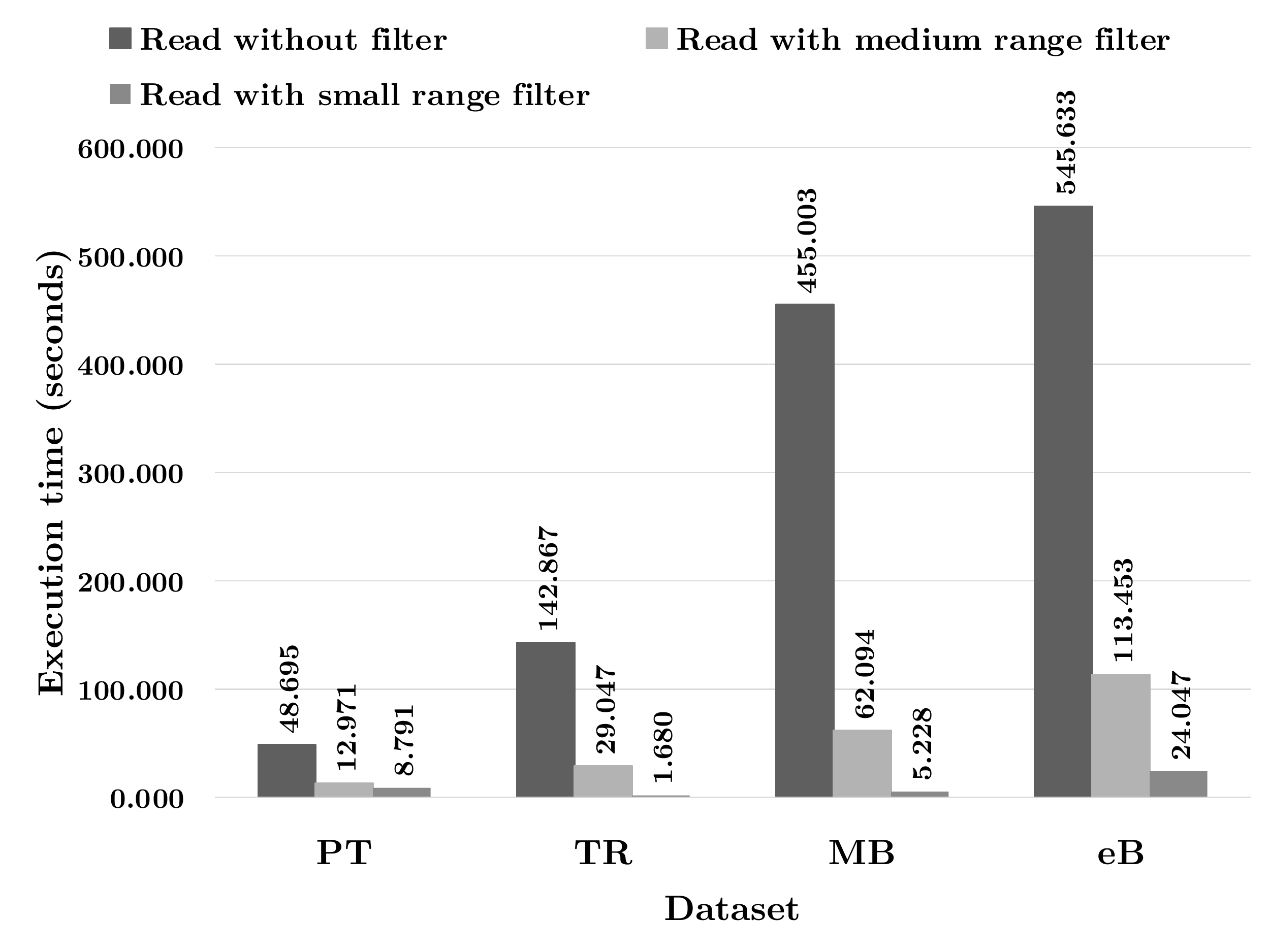}
    \caption{The performance of the light-weight spatial index}
    \label{fig:read_filter}
\end{figure}

Parquet by default collects column statistics for column groups, and chunks. In this experiment, we show the case when no filter is applied, and two additional cases with a small range filter, covering less than 0.01\% of the total area covered by the dataset, and a somewhat larger range filter, covering something between 0.33\% to 4\% depending on the dataset. Figure~\ref{fig:read_filter} shows these results for reading based on these configurations. Note that this filtering is applied per column group first, and then per column chunk. The figure clearly highlights the benefit of this type of filtering. Note that GeoParquet has similar benefit in terms of pruning parts of columns, but it stores additional columns for the minimum-bounding-rectangle and applies the filters based on them. The current default implementation of Parquet doesn't support filters on repeated columns, however, we made a slight modification to its source code to make it work in our case.


This is only the most basic type of pruning which is provided out-of-the-box. However, there are several opportunities for more advanced indexing. Additional metadata can be stored on each column and retrieved when reading prior to unpacking the values. By implementing custom filters specific to spatial indexes, it should be possible to apply more advanced pruning. Additionally, more accurate sorting can be applied by sorting the entire dataset efficiently. The partitioning techniques available in Beast~\cite{eldawy2021beast} can be used to do so efficiently.

\section{Related Work}
\label{sec:related-work}

This section covers the related work in two areas, column formats and encoding.

\paragraph{Column Formats}
Column stores~\cite{SAB+05} have been proposed for data warehousing and analytical queries due to their efficient storage and retrieval. To support semi-structured big-data with nesting and repetition, Dremel~\cite{MGL+10} was introduced by Google which then inspired the open-source Parquet file format~\cite{noauthor_parquet_2022}. It is widely used across many applications for data storage and data analysis. Several big-data systems adopted Parquet as one of its standard formats such as Apache Spark~\cite{zaharia2012fast} which uses it in its Spark SQL~\cite{armbrust2015spark} and MLlib~\cite{meng2016mllib}. An experimental evaluation~\cite{10.14778/2732977.2733002} showed the efficiency of Parquet with text data. We plan to integrate SpatialParquet within existing big spatial data systems, e.g., Beast~\cite{eldawy2021beast}, which would make it possible to benefit from the existing functionality in Spark for the Parquet data format. The only existing attempt to provide a column-oriented format for geo-spatial data is GeoParquet~\cite{geoparquet}, also referred to as geo-arrow, which encodes the geometry value in the Well-Known Binary (WKB) format. However, as shown in the experiments, this does not provide a good output size since it can only apply general purpose compression methods.

\paragraph{Encoding}
Parquet ships with encoding techniques for integer and string values, e.g., delta, run-length, and dictionary encoding\cite{5485951,muller2014adaptive,8316293,10.1145/3329785.3329924}. We use RLE for the type column but none of these techniques work with floating-point coordinates.
Due to the complexity of encoding floating-point values, some recent work proposed methods that are tailored for specific applications, however, none of these focuses on geographic coordinates.
Gorilla~\cite{pelkonen2015gorilla} targets time series data. It applies XOR between consecutive values and adds post-processing steps to remove leading and trailing zeros. Our preliminary investigation determined that this method does not work well for geographic data and we found that the proposed method is more efficient with geographic data. Similarly, the work in~\cite{blalock2018sprintz} focuses on time series data and improves over Gorilla~\cite{pelkonen2015gorilla}. Also, \cite{9005580} focuses on time series data but provides a different approach by encoding similar patterns in time series by mapping them to a dictionary.

Furthermore, other literature focused on the lossless compression of scientific data. The work in~\cite{1607248} uses the integer delta encoding and XOR for encoding the values, as well as, a hash-table for predicting the previous value. Other examples of lossless compression of scientific data include~\cite{6327234,8416606,9418774}. These compression techniques can be applied after the delta encoding but we leave this work for future research. We do not consider lossy compression techniques~\cite{7516088,8421751,9458791,8622520} in this paper since we propose a general-purpose storage format while lossy techniques can be applied only to some applications.

There are several general optimizations that can be added for column structured data. For example, the work in \cite{10.1145/3318464.3384413} explores how re-arranging the data can help improve the space saved with run-length-encoding (RLE), which is something that we have discussed in the experiments. RLE is not specific to any data type. Moreover, \cite{10.1145/3211922.3211932} provides a SIMD based optimization for better filtering of column data. It claims that this optimization can be faster by 90\% than the default filtering in Parquet.

In summary, and to the best of our knowledge, SptialParquet is the only system that proposes a specialized lossless encoding technique for floating-point geographic coordinates.

\section{Conclusion}
\label{sec:conclusion}
This paper introduced SpatialParquet, a column-oriented file format for geospatial data. SpatialParquet is designed to store large-scale spatial data in a column format that reduces disk size and improves the performance of analytical queries. To accomplish its goals, we explained how SpatialParquet introduces a Parquet data type that structures all common geospatial data types, e.g., points, lines, and polygons, in a format that is compatible with Parquet. To make the storage of floating-point coordinate values more efficient, SpatialParquet introduced the FP-delta encoder, which is an efficient encoder that captures the redundancy in geospatial attributes and utilizes it to reduce the storage size. Finally, SpatialParquet used column statistics from Parquet to build a light-weight spatial index that can reduce disk access by skipping file pages that do not overlap with a spatial query range. Experiments on large-scale real data showed that SpatialParquet outperforms all popular spatial file formats when it comes to data analytics.

\bibliographystyle{ACM-Reference-Format}
\bibliography{references}


\begin{thebibliography}{33}


\ifx \showCODEN    \undefined \def \showCODEN     #1{\unskip}     \fi
\ifx \showDOI      \undefined \def \showDOI       #1{#1}\fi
\ifx \showISBNx    \undefined \def \showISBNx     #1{\unskip}     \fi
\ifx \showISBNxiii \undefined \def \showISBNxiii  #1{\unskip}     \fi
\ifx \showISSN     \undefined \def \showISSN      #1{\unskip}     \fi
\ifx \showLCCN     \undefined \def \showLCCN      #1{\unskip}     \fi
\ifx \shownote     \undefined \def \shownote      #1{#1}          \fi
\ifx \showarticletitle \undefined \def \showarticletitle #1{#1}   \fi
\ifx \showURL      \undefined \def \showURL       {\relax}        \fi
\providecommand\bibfield[2]{#2}
\providecommand\bibinfo[2]{#2}
\providecommand\natexlab[1]{#1}
\providecommand\showeprint[2][]{arXiv:#2}

\bibitem[arm(2015)]%
        {armbrust2015spark}
 \bibinfo{year}{2015}\natexlab{}.
\newblock \showarticletitle{{Spark sql: Relational data processing in spark}}.
  In \bibinfo{booktitle}{\emph{SIGMOD}}. \bibinfo{pages}{1383--1394}.
\newblock


\bibitem[noa(2022)]%
        {noauthor_parquet_2022}
 \bibinfo{year}{2022}\natexlab{}.
\newblock \bibinfo{title}{Parquet}.
\newblock
\newblock
\urldef\tempurl%
\url{https://github.com/apache/parquet-format}
\showURL{%
\tempurl}
\newblock
\shownote{original-date: 2014-06-10T07:00:07Z}.


\bibitem[Aghav(2010)]%
        {5485951}
\bibfield{author}{\bibinfo{person}{Sushila Aghav}.}
  \bibinfo{year}{2010}\natexlab{}.
\newblock \showarticletitle{Database compression techniques for performance
  optimization}. In \bibinfo{booktitle}{\emph{2010 2nd International Conference
  on Computer Engineering and Technology}}, Vol.~\bibinfo{volume}{6}.
  \bibinfo{pages}{V6--714--V6--717}.
\newblock
\urldef\tempurl%
\url{https://doi.org/10.1109/ICCET.2010.5485951}
\showDOI{\tempurl}


\bibitem[Al{-}Mamun et~al\mbox{.}(2020)]%
        {AWA20}
\bibfield{author}{\bibinfo{person}{Abdullah Al{-}Mamun}, \bibinfo{person}{Hao
  Wu}, {and} \bibinfo{person}{Walid~G. Aref}.} \bibinfo{year}{2020}\natexlab{}.
\newblock \showarticletitle{A Tutorial on Learned Multi-dimensional Indexes}.
  In \bibinfo{booktitle}{\emph{SIGSPATIAL}}. \bibinfo{publisher}{{ACM}},
  \bibinfo{address}{Seattle, WA}, \bibinfo{pages}{1--4}.
\newblock


\bibitem[Austin et~al\mbox{.}(2016)]%
        {7516088}
\bibfield{author}{\bibinfo{person}{Woody Austin}, \bibinfo{person}{Grey
  Ballard}, {and} \bibinfo{person}{Tamara~G. Kolda}.}
  \bibinfo{year}{2016}\natexlab{}.
\newblock \showarticletitle{Parallel Tensor Compression for Large-Scale
  Scientific Data}. In \bibinfo{booktitle}{\emph{IPDPS}}.
  \bibinfo{pages}{912--922}.
\newblock


\bibitem[Blalock et~al\mbox{.}(2018)]%
        {blalock2018sprintz}
\bibfield{author}{\bibinfo{person}{Davis Blalock}, \bibinfo{person}{Samuel
  Madden}, {and} \bibinfo{person}{John Guttag}.}
  \bibinfo{year}{2018}\natexlab{}.
\newblock \showarticletitle{Sprintz: Time series compression for the internet
  of things}.
\newblock \bibinfo{journal}{\emph{Proceedings of the ACM on Interactive,
  Mobile, Wearable and Ubiquitous Technologies}} \bibinfo{volume}{2},
  \bibinfo{number}{3} (\bibinfo{year}{2018}), \bibinfo{pages}{1--23}.
\newblock


\bibitem[Claggett et~al\mbox{.}(2018)]%
        {8416606}
\bibfield{author}{\bibinfo{person}{Steven Claggett}, \bibinfo{person}{Sahar
  Azimi}, {and} \bibinfo{person}{Martin Burtscher}.}
  \bibinfo{year}{2018}\natexlab{}.
\newblock \showarticletitle{SPDP: An Automatically Synthesized Lossless
  Compression Algorithm for Floating-Point Data}. In
  \bibinfo{booktitle}{\emph{2018 Data Compression Conference}}.
  \bibinfo{pages}{335--344}.
\newblock


\bibitem[data.gov(2022)]%
        {datagov}
data.gov \bibinfo{year}{2022}\natexlab{}.
\newblock \bibinfo{title}{{Data.Gov: The home of the U.S. Government's open
  data}}.
\newblock
\newblock


\bibitem[Deutsch(1996)]%
        {deutsch1996gzip}
\bibfield{author}{\bibinfo{person}{Peter Deutsch}.}
  \bibinfo{year}{1996}\natexlab{}.
\newblock \bibinfo{booktitle}{\emph{GZIP file format specification version
  4.3}}.
\newblock \bibinfo{type}{{T}echnical {R}eport}.
\newblock


\bibitem[Di et~al\mbox{.}(2019)]%
        {8421751}
\bibfield{author}{\bibinfo{person}{Sheng Di}, \bibinfo{person}{Dingwen Tao},
  \bibinfo{person}{Xin Liang}, {and} \bibinfo{person}{Franck Cappello}.}
  \bibinfo{year}{2019}\natexlab{}.
\newblock \showarticletitle{Efficient Lossy Compression for Scientific Data
  Based on Pointwise Relative Error Bound}.
\newblock \bibinfo{journal}{\emph{TPDS}} \bibinfo{volume}{30},
  \bibinfo{number}{2} (\bibinfo{year}{2019}), \bibinfo{pages}{331--345}.
\newblock


\bibitem[Eldawy et~al\mbox{.}(2015)]%
        {EAM15}
\bibfield{author}{\bibinfo{person}{Ahmed Eldawy}, \bibinfo{person}{Louai
  Alarabi}, {and} \bibinfo{person}{Mohamed~F. Mokbel}.}
  \bibinfo{year}{2015}\natexlab{}.
\newblock \showarticletitle{{Spatial Partitioning Techniques in Spatial
  Hadoop}}.
\newblock \bibinfo{journal}{\emph{{PVLDB}}} \bibinfo{volume}{8},
  \bibinfo{number}{12} (\bibinfo{year}{2015}), \bibinfo{pages}{1602--1605}.
\newblock


\bibitem[Eldawy et~al\mbox{.}(2021)]%
        {eldawy2021beast}
\bibfield{author}{\bibinfo{person}{Ahmed Eldawy}, \bibinfo{person}{Vagelis
  Hristidis}, \bibinfo{person}{Saheli Ghosh}, \bibinfo{person}{Majid Saeedan},
  \bibinfo{person}{Akil Sevim}, \bibinfo{person}{AB Siddique},
  \bibinfo{person}{Samriddhi Singla}, \bibinfo{person}{Ganesh Sivaram},
  \bibinfo{person}{Tin Vu}, {and} \bibinfo{person}{Yaming Zhang}.}
  \bibinfo{year}{2021}\natexlab{}.
\newblock \showarticletitle{Beast: Scalable Exploratory Analytics on
  Spatio-temporal Data}. In \bibinfo{booktitle}{\emph{CIKM}}.
  \bibinfo{pages}{3796--3807}.
\newblock


\bibitem[Floratou et~al\mbox{.}(2014)]%
        {10.14778/2732977.2733002}
\bibfield{author}{\bibinfo{person}{Avrilia Floratou},
  \bibinfo{person}{Umar~Farooq Minhas}, {and} \bibinfo{person}{Fatma
  \"{O}zcan}.} \bibinfo{year}{2014}\natexlab{}.
\newblock \showarticletitle{SQL-on-Hadoop: Full Circle Back to Shared-Nothing
  Database Architectures}.
\newblock \bibinfo{journal}{\emph{PVLDB}} \bibinfo{volume}{7},
  \bibinfo{number}{12} (\bibinfo{year}{2014}).
\newblock


\bibitem[Fout and Ma(2012)]%
        {6327234}
\bibfield{author}{\bibinfo{person}{Nathaniel Fout} {and}
  \bibinfo{person}{Kwan-Liu Ma}.} \bibinfo{year}{2012}\natexlab{}.
\newblock \showarticletitle{An Adaptive Prediction-Based Approach to Lossless
  Compression of Floating-Point Volume Data}.
\newblock \bibinfo{journal}{\emph{TVCG}} \bibinfo{volume}{18},
  \bibinfo{number}{12} (\bibinfo{year}{2012}), \bibinfo{pages}{2295--2304}.
\newblock


\bibitem[geoparquet(2022)]%
        {geoparquet}
geoparquet \bibinfo{year}{2022}\natexlab{}.
\newblock \bibinfo{title}{{GeoParquet: Store Vector Data in Apache Parquet}}.
\newblock
\newblock
\urldef\tempurl%
\url{https://github.com/opengeospatial/geoparquet}
\showURL{%
\tempurl}


\bibitem[Ghosh et~al\mbox{.}(2019)]%
        {GVE+19}
\bibfield{author}{\bibinfo{person}{Saheli Ghosh} {et~al\mbox{.}}}
  \bibinfo{year}{2019}\natexlab{}.
\newblock \showarticletitle{{UCR-STAR: The UCR Spatio-Temporal Active
  Repository}}.
\newblock \bibinfo{journal}{\emph{SIGSPATIAL Special}} \bibinfo{volume}{11},
  \bibinfo{number}{2} (\bibinfo{date}{Dec.} \bibinfo{year}{2019}),
  \bibinfo{pages}{34–40}.
\newblock


\bibitem[Jiang and Elmore(2018)]%
        {10.1145/3211922.3211932}
\bibfield{author}{\bibinfo{person}{Hao Jiang} {and} \bibinfo{person}{Aaron~J.
  Elmore}.} \bibinfo{year}{2018}\natexlab{}.
\newblock \showarticletitle{Boosting Data Filtering on Columnar Encoding with
  SIMD}. In \bibinfo{booktitle}{\emph{DaMoN@SIGMOD}}. Article
  \bibinfo{articleno}{6}, \bibinfo{numpages}{10}~pages.
\newblock
\showISBNx{9781450358538}


\bibitem[Kanda et~al\mbox{.}(2017)]%
        {8316293}
\bibfield{author}{\bibinfo{person}{Shunsuke Kanda}, \bibinfo{person}{Kazuhiro
  Morita}, {and} \bibinfo{person}{Masao Fuketa}.}
  \bibinfo{year}{2017}\natexlab{}.
\newblock \showarticletitle{Practical String Dictionary Compression Using
  String Dictionary Encoding}. In \bibinfo{booktitle}{\emph{2017 International
  Conference on Big Data Innovations and Applications (Innovate-Data)}}.
  \bibinfo{pages}{1--8}.
\newblock


\bibitem[Khelifati et~al\mbox{.}(2019)]%
        {9005580}
\bibfield{author}{\bibinfo{person}{Abdelouahab Khelifati},
  \bibinfo{person}{Mourad Khayati}, {and} \bibinfo{person}{Philippe
  Cudré-Mauroux}.} \bibinfo{year}{2019}\natexlab{}.
\newblock \showarticletitle{CORAD: Correlation-Aware Compression of Massive
  Time Series using Sparse Dictionary Coding}. In
  \bibinfo{booktitle}{\emph{IEEE BigData}}. \bibinfo{pages}{2289--2298}.
\newblock


\bibitem[Knorr et~al\mbox{.}(2021)]%
        {9418774}
\bibfield{author}{\bibinfo{person}{Fabian Knorr}, \bibinfo{person}{Peter
  Thoman}, {and} \bibinfo{person}{Thomas Fahringer}.}
  \bibinfo{year}{2021}\natexlab{}.
\newblock \showarticletitle{ndzip: {A} High-Throughput Parallel Lossless
  Compressor for Scientific Data}. In \bibinfo{booktitle}{\emph{DCC}}.
  \bibinfo{pages}{103--112}.
\newblock


\bibitem[Lasch et~al\mbox{.}(2019)]%
        {10.1145/3329785.3329924}
\bibfield{author}{\bibinfo{person}{Robert Lasch} {et~al\mbox{.}}}
  \bibinfo{year}{2019}\natexlab{}.
\newblock \showarticletitle{Fast \& Strong: The Case of Compressed String
  Dictionaries on Modern CPUs}. In \bibinfo{booktitle}{\emph{DaMoN@SIGMOD}}.
  Article \bibinfo{articleno}{4}, \bibinfo{numpages}{10}~pages.
\newblock
\showISBNx{9781450368018}


\bibitem[Liang et~al\mbox{.}(2018)]%
        {8622520}
\bibfield{author}{\bibinfo{person}{Xin Liang} {et~al\mbox{.}}}
  \bibinfo{year}{2018}\natexlab{}.
\newblock \showarticletitle{Error-Controlled Lossy Compression Optimized for
  High Compression Ratios of Scientific Datasets}. In
  \bibinfo{booktitle}{\emph{IEEE BigData}}. \bibinfo{pages}{438--447}.
\newblock


\bibitem[Melnik et~al\mbox{.}(2010)]%
        {MGL+10}
\bibfield{author}{\bibinfo{person}{Sergey Melnik} {et~al\mbox{.}}}
  \bibinfo{year}{2010}\natexlab{}.
\newblock \showarticletitle{Dremel: Interactive Analysis of Web-Scale
  Datasets}.
\newblock \bibinfo{journal}{\emph{PVLDB}} \bibinfo{volume}{3},
  \bibinfo{number}{1} (\bibinfo{year}{2010}), \bibinfo{pages}{330--339}.
\newblock


\bibitem[Meng et~al\mbox{.}(2016)]%
        {meng2016mllib}
\bibfield{author}{\bibinfo{person}{Xiangrui Meng} {et~al\mbox{.}}}
  \bibinfo{year}{2016}\natexlab{}.
\newblock \showarticletitle{Mllib: Machine learning in apache spark}.
\newblock \bibinfo{journal}{\emph{The Journal of Machine Learning Research}}
  \bibinfo{volume}{17}, \bibinfo{number}{1} (\bibinfo{year}{2016}),
  \bibinfo{pages}{1235--1241}.
\newblock


\bibitem[M{\"u}ller et~al\mbox{.}(2014)]%
        {muller2014adaptive}
\bibfield{author}{\bibinfo{person}{Ingo M{\"u}ller}, \bibinfo{person}{Cornelius
  Ratsch}, \bibinfo{person}{Franz Faerber}, {et~al\mbox{.}}}
  \bibinfo{year}{2014}\natexlab{}.
\newblock \showarticletitle{Adaptive String Dictionary Compression in In-Memory
  Column-Store Database Systems.}. In \bibinfo{booktitle}{\emph{EDBT}},
  Vol.~\bibinfo{volume}{14}. \bibinfo{pages}{283--294}.
\newblock


\bibitem[Pelkonen et~al\mbox{.}(2015)]%
        {pelkonen2015gorilla}
\bibfield{author}{\bibinfo{person}{Tuomas Pelkonen} {et~al\mbox{.}}}
  \bibinfo{year}{2015}\natexlab{}.
\newblock \showarticletitle{Gorilla: A fast, scalable, in-memory time series
  database}.
\newblock \bibinfo{journal}{\emph{PVLDB}} \bibinfo{volume}{8},
  \bibinfo{number}{12} (\bibinfo{year}{2015}), \bibinfo{pages}{1816--1827}.
\newblock


\bibitem[Ratanaworabhan et~al\mbox{.}(2006)]%
        {1607248}
\bibfield{author}{\bibinfo{person}{P. Ratanaworabhan}, \bibinfo{person}{Jian
  Ke}, {and} \bibinfo{person}{M. Burtscher}.} \bibinfo{year}{2006}\natexlab{}.
\newblock \showarticletitle{Fast lossless compression of scientific
  floating-point data}. In \bibinfo{booktitle}{\emph{DCC}}.
  \bibinfo{pages}{133--142}.
\newblock


\bibitem[Shi(2020)]%
        {10.1145/3318464.3384413}
\bibfield{author}{\bibinfo{person}{Jia Shi}.} \bibinfo{year}{2020}\natexlab{}.
\newblock \showarticletitle{Column Partition and Permutation for Run Length
  Encoding in Columnar Databases}. In \bibinfo{booktitle}{\emph{SIGMOD}}.
  \bibinfo{publisher}{Association for Computing Machinery},
  \bibinfo{pages}{2873–2874}.
\newblock
\showISBNx{9781450367356}


\bibitem[Stonebraker et~al\mbox{.}(2005)]%
        {SAB+05}
\bibfield{author}{\bibinfo{person}{Michael Stonebraker} {et~al\mbox{.}}}
  \bibinfo{year}{2005}\natexlab{}.
\newblock \showarticletitle{C-Store: {A} Column-oriented {DBMS}}. In
  \bibinfo{booktitle}{\emph{VLDB}}. \bibinfo{pages}{553--564}.
\newblock


\bibitem[Vohra(2016)]%
        {V16}
\bibfield{author}{\bibinfo{person}{Deepak Vohra}.}
  \bibinfo{year}{2016}\natexlab{}.
\newblock \bibinfo{booktitle}{\emph{Apache Parquet}}.
\newblock \bibinfo{publisher}{Apress}, \bibinfo{address}{Berkeley, CA},
  \bibinfo{pages}{325--335}.
\newblock
\showISBNx{978-1-4842-2199-0}
\urldef\tempurl%
\url{https://doi.org/10.1007/978-1-4842-2199-0_8}
\showDOI{\tempurl}


\bibitem[Vu and Eldawy(2020)]%
        {VE20b}
\bibfield{author}{\bibinfo{person}{Tin Vu} {and} \bibinfo{person}{Ahmed
  Eldawy}.} \bibinfo{year}{2020}\natexlab{}.
\newblock \showarticletitle{{R*-Grove: Balanced Spatial Partitioning for
  Large-Scale Datasets}}.
\newblock  (\bibinfo{date}{Aug.} \bibinfo{year}{2020}).
\newblock
\urldef\tempurl%
\url{https://doi.org/10.3389/fdata.2020.00028}
\showDOI{\tempurl}


\bibitem[Zaharia et~al\mbox{.}(2012)]%
        {zaharia2012fast}
\bibfield{author}{\bibinfo{person}{Matei Zaharia} {et~al\mbox{.}}}
  \bibinfo{year}{2012}\natexlab{}.
\newblock \showarticletitle{Fast and interactive analytics over Hadoop data
  with Spark}.
\newblock \bibinfo{journal}{\emph{Usenix Login}} \bibinfo{volume}{37},
  \bibinfo{number}{4} (\bibinfo{year}{2012}), \bibinfo{pages}{45--51}.
\newblock


\bibitem[Zhao et~al\mbox{.}(2021)]%
        {9458791}
\bibfield{author}{\bibinfo{person}{Kai Zhao} {et~al\mbox{.}}}
  \bibinfo{year}{2021}\natexlab{}.
\newblock \showarticletitle{Optimizing Error-Bounded Lossy Compression for
  Scientific Data by Dynamic Spline Interpolation}. In
  \bibinfo{booktitle}{\emph{ICDE}}. \bibinfo{pages}{1643--1654}.
\newblock


\end{thebibliography}

\appendix

\section{Preliminaries}
\label{sec:preliminaries}

This appendix provides some preliminaries that are needed to understand this paper in case the reader needs a quick memory refresh.

\subsection{Geometry Data Types}
Vector geometry data is represented as points, lines, and polygons. The Open Geospatial Consortium (OGC) defines an industry standard for representing geospatial data. It defines primarily seven data types as detailed shortly. Notice that each of these data types has a variation that stores three-dimensional points $(x,y,z)$ and four dimensional points $(x,y,z,m)$. We focus on two-dimensional points for simplicity but all the proposed techniques in this paper can seamlessly apply for any number of dimensions.
\begin{enumerate}
    \item {\bf Point} is defined by a single coordinate $(x,y)$.
    \item {\bf LineString} is defined as an ordered sequence of coordinates. A special case of LineString is a {\em Ring} which has the same starting and ending points.
    \item {\bf Polygon} is defined as a sequence of Rings. The first Ring defines the outer shell and subsequent rings define holes in the polygon.
    \item {\bf MultiPoint} is a set of Points.
    \item {\bf MultiLineString} is a set of LineStrings.
    \item {\bf MultiPolygon} is a set of Polygons.
    \item {\bf GeometryCollection} is a set of geometries that can include nested GeometryCollections.
\end{enumerate}

\subsection{Parquet}
Parquet is a column-oriented file format that is geared for big data storage. It can store any data type that can be defined using Google Protocol Buffers Format (PBF). Basically, PBF can store primitive values, e.g., numbers and strings, arrays, and nested objects. Parquet only stores the primitive values in columns. Nesting and repetition are supported by attaching definition and repetition levels to each column as further detailed in~\cite{MGL+10}. The only limitation in Parquet is that all values in the file have to follow the same exact PBF schema. Therefore, two records cannot contain mismatching values for the same column, e.g., string and number.

Parquet groups the records into {\em row groups} which are typically 1GB in size. Within each group, each column is stored separately as a {\em column chunk}. Each column chunk is further split into {\em pages}. Finally, each page is encoded and compressed to improve storage efficiency. When reading a file back, the minimum reading unit is a page. By default, each page is about 1MB of size.

Parquet provides the delta encoding and run-length-encoding (RLE) to reduce the storage size for each page. It also provides lossless compression techniques such as GZip and Snappy to reduce the storage size. Finally, Parquet can collect statistics for each page, e.g., minimum and maximum, and can use this information to skip reading pages that do not fall within a user-provided query range. For example, if a user wants to read all entries with the anomalous temperature of $125^\circ F$ or more, Parquet can skip all pages that have degrees below $125$.

\subsection{IEEE Floating Point Format}
In computers, floating-point values are represented in the IEEE 754 floating point standard as shown in Figure~\ref{fig:ieee-754}. The value is stored in three parts, sign, exponent, and fraction. The value stored in this format can be calculated as:
\begin{equation}
(-1)^{sign}(1.fraction)_2 \times 2^{exponent-1023}
\end{equation}

\begin{figure}[h]
    \centering
    \begin{tabular}{|p{0.3in}|p{0.75in}|p{1.5in}|}
    \hline
    Sign & Exponent & Fraction \\
    1-bit & 11 bits & 52 bits \\
    \hline
    \end{tabular}
    \caption{The IEEE 754 standard for floating point numbers}
    \label{fig:ieee-754}
\end{figure}

The key idea behind this representation is that any binary floating-point number can be represented in the scientific notation as $1.frac\times 2^{exp}$. Therefore, any value is first normalized to make the {\em decimal point}\footnote{We use the term {\em decimal point} for convenience even when describing binary numbers.} right after the most-significant one bit. After that, only the fraction and exponent need to be stored.

\section{Algorithm Pseudo-Codes}
\label{sec:pseudo-code}

\begin{algorithm}[t]
\begin{algorithmic}[1]
\Function{computeBestDeltaBits}{double[] $X$}
\State $h$ = Array[0..64] \Comment The histogram has 65 bins
\For{$i$ = 1 {\bf to} $|X|-1$}\label{alg:best-n:for-x-start}
  \State delta = cast-long($X[i]$)-cast-long($X[i-1]$)
  \State zigzag = (delta $\gg$ 63) $\oplus$ (delta $\ll$ 1)
  \State $n$ = num-significant-bits(zigzag)
  \State $h[n]++$
\EndFor \label{alg:best-n:for-x-end}
\For{$n$ = 63 {\bf downto} 0} \Comment{Compute suffix sum} \label{alg:best-n:suffix-sum}
  \State $h[n]+=h[n+1]$
\EndFor
\State $n^*$ = 0
\State $S_{min} = 64(|X|-1)$
\For{$n$ = 1 {\bf to} 63}
  \State $S=n\cdot (|X|-1)+64\cdot h[n]$
  \If{$S<S_{min}$}
    \State $(n^*,S_{min}) = (n, S)$
  \EndIf
\EndFor
\State \Return $n^*$
\EndFunction
\end{algorithmic}
\caption{Find the number of bits to minimize the output size}
\label{alg:best-n}
\end{algorithm}

Algorithm~\ref{alg:best-n} provides the pseudo-code for the algorithm that computes the number of bits that minimizes the output size. First it initializes the histogram that captures the number of deltas for each number of bits. Notice that the histogram has 65 bins since the number of bits can go from zero to 65. The for loop in Lines~\ref{alg:best-n:for-x-start}-\label{alg:best-n:for-x-end}, scans all the values and computes the delta as done in the original algorithm. For each value, it computes the minimum number of bits required for this delta and increments the corresponding bin in the histogram. After that, the loop in Line~\ref{alg:best-n:suffix-sum} calculates the {\em suffix sum} of the histogram. This is to allow the summation in Equation~\ref{eqn:outputsize} to be calculated in constant time. After that, to find the best value, we simply try all the 65 possible values and choose the one that yields the minimum output size by applying Equation~\ref{eqn:outputsize}. Finally, we return the best value $n^*$ that corresponds to the minimum.

\end{document}